\documentclass[12pt]{article}
\pdfoutput=1

\usepackage{pgfplots}
\usetikzlibrary{decorations.pathmorphing}

\tikzset{snake it/.style={decorate, decoration=snake}}
\pgfplotsset{compat=1.10}
\usepgfplotslibrary{fillbetween}
\usetikzlibrary{patterns}

\usepackage{draft} 
\usepackage{hyperref}
\usepackage{graphicx,color,subfig}
\usepackage{cite}
\usepackage{mciteplus}
\usepackage{skak}
\usepackage{empheq}
\usepackage{tikz}
\usepackage{bbm}
\usepackage{makecell}

\usepackage{bm}
\usepackage{braket}

\DeclareFontFamily{OT1}{pzc}{}
\DeclareFontShape{OT1}{pzc}{m}{it}{<-> s * [1.10] pzcmi7t}{}
\DeclareMathAlphabet{\mathpzc}{OT1}{pzc}{m}{it}

\usetikzlibrary{snakes}
\usetikzlibrary{arrows}
\usetikzlibrary{decorations.pathmorphing}
\tikzset{snake it/.style={decorate, decoration=snake}}
\usetikzlibrary{shapes.misc}
\tikzset{cross/.style={cross out, draw=black, minimum size=2*(#1-\pgflinewidth), inner sep=0pt, outer sep=0pt},
cross/.default={1pt}}

\def\be#1\ee{\begin{align}#1\end{align}}

\makeatletter
\newcommand{\bdryno}{\mathpalette\bdry@no\relax}
\newcommand{\bdry@no}[2]{%
  \mspace{1mu}%
  \vbox{%
    \hbox{$\m@th#1\scriptstyle{\ast}$}
    \nointerlineskip
    \kern.25ex
    \hbox{$\m@th#1\scriptstyle{\ast}$}
    \kern-.06ex
  }%
  \mspace{1mu}%
}
\makeatother

\usetikzlibrary{shapes.misc}

\usetikzlibrary{decorations.markings}
\usetikzlibrary{shapes.misc}

\tikzset{cross/.style={cross out, draw=black, minimum size=2*(#1-\pgflinewidth), inner sep=0pt, outer sep=0pt},
cross/.default={1pt}}
    
\tikzset{
   mydisk/.pic={
        \filldraw[color=black, fill=black!30, thick] circle (0.25);
        \node[color=black] at (0.0,0.45) {\tiny $#1$};
        \draw node[cross=3pt, very thick] {};
    }}
    
\tikzset{
  pics/disk2/.style n args={2}{
    code = { %
        \filldraw[color=black, fill=black!30, thick] circle (0.3);
        \node[color=black] at (0,0.5) {\scriptsize ${#1}_{#2}$};
        \draw node[cross=3pt, very thick] {};
    }
  }
}
\tikzset{
  pics/disk3/.style n args={3}{
    code = { %
        \filldraw[color=black, fill=black!30, thick] circle (0.3);
        \node[color=black] at (0,0.5) {\scriptsize ${#1}_{#2}$};
        \draw node[cross=3pt, very thick] {};
        \node[color=black] at (-0.02,-0.17) {\scriptsize $#3$};
    }
  }
}

\tikzset{
  pics/diskbig/.style n args={2}{
    code = { %
        \filldraw[color=black, fill=black!30, thick] circle (0.55);
        \node[color=black] at (0,0.75) {\scriptsize $#1$};
        \draw node[cross=3pt, very thick] {};
        \node[color=black] at (0,0.2) {\small $#2$};
    }
  }
}

\tikzset{
  pics/disk1/.style n args={1}{
    code = { %
        \filldraw[color=black, fill=black!30, thick] circle (0.3);
        \node[color=black] at (0,0.5) {\scriptsize ${#1}$};
        \draw node[cross=3pt, very thick] {};
    }
  }
}   
\tikzset{
  pics/diskn/.style n args={1}{
    code = { %
        \filldraw[color=black, densely dashed, fill=black!30, thick] circle (0.3);
        \node[color=black] at (0,0.5) {\scriptsize ${#1}$};
        \draw node[cross=3pt, very thick] {};
    }
  }
}    
\tikzset{
  pics/disk1w/.style n args={2}{
    code = { %
        \filldraw[color=black, fill=black!30, thick] circle (0.35);
        \node[color=black] at (0,0.5) {\scriptsize ${#1}$};
        \draw node[cross=3pt, very thick] {};
        \node[color=black] at (-0.02,-0.17) {\scriptsize ${#2}$};
    }
  }
}
    
\tikzset{
  pics/cyl/.style n args={2}{
    code = { %
        \filldraw[color=black, fill=black!30, thick] circle (0.3);
        \filldraw[color=black, fill=white, thick] circle (0.15);
        \node[color=black] at (0.05,0.0) {\tiny $#1$};
        \node[color=black] at (0.4,0.0) {\tiny $#2$};
    }
  }
}

\tikzset{
  pics/cyl2/.style n args={2}{
    code = { %
        \filldraw[color=black, fill=black!30, thick] circle (0.5);
        \filldraw[color=black, fill=white, thick]  circle (0.25);
        \node[color=black] at (0.08,0) {\small $#1$};
        \node[color=black] at (0.67,0) {\small $#2$};

    }
  }
}
\tikzset{
  pics/cyl3/.style n args={4}{
    code = { %
        \filldraw[color=black, fill=black!30, thick] circle (0.35);
        \filldraw[color=black, fill=white, thick]  circle (0.22);
        \node[color=black] at (0,0.51) {\scriptsize ${#1}_{#2}$};
        \node[color=black] at (0,0) {\scriptsize ${#3}_{#4}$};

    }
  }
}

\tikzset{
  pics/cyl11/.style n args={2}{
    code = { %
        \filldraw[color=black, fill=black!30, thick] circle (0.35);
        \filldraw[color=black, fill=white, thick]  circle (0.22);
        \node[color=black] at (0,0.51) {\scriptsize ${#1}$};
        \node[color=black] at (0,0) {\scriptsize ${#2}$};

    }
  }
}

\tikzset{
  pics/cyl11/.style n args={2}{
    code = { %
        \filldraw[color=black, fill=black!30, thick] circle (0.35);
        \filldraw[color=black, fill=white, thick]  circle (0.17);
        \node[color=black] at (0,0.51) {\scriptsize ${#1}$};
        \node[color=black] at (0,0) {\scriptsize ${#2}$};

    }
  }
}

\tikzset{
  pics/cyln1/.style n args={2}{
    code = { %
        \filldraw[color=black, densely dashed,fill=black!30, thick] circle (0.35);
        \filldraw[color=black, fill=white, thick]  circle (0.17);
        \node[color=black] at (0,0.51) {\scriptsize ${#1}$};
        \node[color=black] at (0,0) {\scriptsize ${#2}$};

    }
  }
}

\tikzset{
  pics/cylnn/.style n args={2}{
    code = { %
        \filldraw[color=black, densely dashed,fill=black!30, thick] circle (0.35);
        \filldraw[color=black, densely dashed, fill=white, thick]  circle (0.17);
        \node[color=black] at (0,0.51) {\scriptsize ${#1}$};
        \node[color=black] at (0,0) {\scriptsize ${#2}$};

    }
  }
}

\begin{document}

\unitlength = .8mm

\begin{titlepage}

\begin{center}

\hfill \\
\hfill \\
\vskip 1cm

\title{Multi-Instanton Calculus in $c=1$ String Theory}

\author{Bruno Balthazar, Victor A. Rodriguez, Xi Yin}

\address{
Jefferson Physical Laboratory, Harvard University, \\
Cambridge, MA 02138 USA
}

\email{bbalthazar@g.harvard.edu, victorrodriguez@g.harvard.edu, xiyin@fas.harvard.edu}

\end{center}

\abstract{ We formulate a strategy for computing the complete set of non-perturbative corrections to closed string scattering in $c=1$ string theory from the worldsheet perspective. This requires taking into account the effect of multiple ZZ-instantons, including higher instantons constructed from ZZ boundary conditions of type $(m,1)$, with a careful treatment of the measure and contour in the integration over the instanton moduli space. The only a priori ambiguity in our prescription is a normalization constant ${\cal N}_{m}$ that appears in the integration measure for the $(m,1)$-type ZZ instanton, at each positive integer $m$. We investigate leading corrections to the closed string reflection amplitude at the $n$-instanton level, i.e. of order $e^{-n/g_s}$, and find striking agreement with our recent proposal on the non-perturbative completion of the dual matrix quantum mechanics, which in turn fixes ${\cal N}_{m}$ for all $m$.  }

\vfill

\end{titlepage}

\eject

\begingroup
\hypersetup{linkcolor=black}
\tableofcontents
\endgroup

\section{Introduction} 

In a recent paper \cite{Balthazar:2019rnh} we proposed that the perturbative closed string scattering amplitudes of $c=1$ string theory, after Borel resummation, should be further corrected by the effect of ZZ-instantons, and the resulting string theory is dual to a natural completion of the $c=1$ matrix quantum mechanics where the fermi sea states with no incoming flux from the ``other side" of the potential are filled. Nontrivial evidence was given at 1-instanton level, at order $e^{-1/g_s}$ for closed string $1\to k$ amplitudes and order $e^{-1/g_s} g_s$ for closed string $1\to 1$ amplitudes, where the string theoretic computation based on worldsheets with boundary on the ZZ-instantons is shown to agree with the proposed matrix model dual, up to the overall normalization factor of the instanton measure, a 2-loop renormalization constant of the instanton action, and a constant in canceling logarithmic divergences between worldsheet diagrams of different topologies. The latter constant was subsequently determined analytically by Sen \cite{Sen:2019qqg} by consideration of open+closed string field theory, in agreement with the numerical result anticipated from \cite{Balthazar:2019rnh}.

In this paper, we extend the analysis of \cite{Balthazar:2019rnh} to multi-instanton levels, namely at order $e^{-n/g_s}$ for all positive integer $n$. Recall that the ZZ-instanton, considered in \cite{Balthazar:2019rnh}, amounts to introducing boundaries on the worldsheet that obey the $(1,1)$ ZZ boundary condition in the Liouville sector and Dirichlet boundary condition in the time-like free boson $X^0$. A class of multi-instanton configurations involve $n$ ZZ-instantons, of the $(1,1)$ type, located at separate Euclidean times $x_1, \cdots, x_n$. We will see that, in addition, one must also consider ``higher instantons" defined by ZZ boundary condition of type $(m,1)$, for $m\geq 2$, whose action is $m$ times that of a single $(1,1)$ ZZ-instanton. For reasons not fully understood, it appears that more general ZZ boundary conditions of type $(m,\ell)$ with $m,\ell\geq 2$ do not contribute. Thus, a general instanton configuration at order $e^{-n/g_s}$ consists of a set of ZZ-instantons of type $(m_i,1)$ located at time $x_i$, for $i=1,\cdots, \ell$, with $\sum_i m_i = n$. We will refer to such an instanton configuration as of type $\{m_1, m_2, \cdots, m_\ell\}$.

\begin{figure}[h]
\centering
\begin{tikzpicture}
\filldraw[color=black, fill=black!30, thick] (0,0) circle (1.25);
\draw (0,0) node[cross=4pt, very thick] {};
\draw (0,0) node[above] {$\cV^{+}_{\omega}$};
\draw[fill] (1.5,0) circle [radius=0.025];
\filldraw[color=black, fill=black!30, thick] (3,0) circle (1.25);
\draw (3,0) node[cross=4pt, very thick] {};
\draw (3,0) node[above] {$\cV^{-}_{\omega_1}$};
\draw[fill] (4.5,0) circle [radius=0.025];
\node at (5.25,0) {$\dots$};
\draw[fill] (6,0) circle [radius=0.025];
\filldraw[color=black, fill=black!30, thick] (7.5,0) circle (1.25);
\draw (7.5,0) node[cross=4pt, very thick] {};
\draw (7.5,0) node[above] {$\cV^{-}_{\omega_k}$};
\end{tikzpicture}
\caption{Worldsheet diagrams that compute the leading non-perturbative $n$-instanton correction to the closed string $1\to k$ amplitude, of order $e^{-n/g_s}$, mediated by multiple ZZ-instantons. For a given configuration set $\{m_1,m_2,\cdots,m_l\}$ with $\Sigma_i m_i = n$, the boundary condition on each disc is given by (\ref{eq:BCdisk}).}
\label{fig:LOninst}
\end{figure}
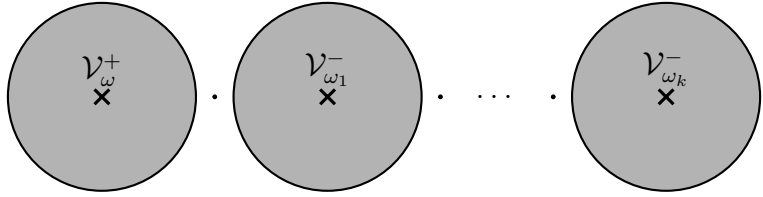

The leading $n$-instanton contribution to $1\to k$ closed string scattering involves worldsheet diagrams that consist of $k+1$ disconnected discs, with one closed string vertex operator inserted on each disc as shown in Figure \ref{fig:LOninst}. The boundary condition is captured by the boundary state
\ie
|B\rangle\rangle = \sum_{i=1}^\ell |{\rm ZZ} (m_i,1)\rangle\rangle_{\rm Liouville} \otimes |D(x_i)\rangle\rangle_{X^0},
\label{eq:BCdisk}
\fe
where $|{\rm ZZ}(m,1)\rangle\rangle_{\rm Liouville}$ refers to the $(m,1)$ ZZ boundary state in Liouville theory and $|D(x)\rangle\rangle_{X^0}$ refers to the Dirichlet boundary state at $X^0 = x$ in the free boson CFT. One must then integrate over the collective coordinates $x_i$, and finally sum over instanton types $\{m_1, \cdots, m_\ell\}$. 

The key nontrivial ingredient will be the choice of contour and measure in the integration over collective coordinates. Let us illustrate our strategy and prescription in the $\{1,1,\cdots, 1\}$ case, i.e.  $n$ instantons of type $(1,1)$. The collective mode integration takes the schematic form
\ie\label{schminst}
e^{-\frac{n}{g_s}} \int \prod_{i=1}^n dx_i\, \mu(x_1,\cdots, x_n) \times \left( {\rm worldsheet~diagram} \right),
\fe
where the worldsheet diagram does not contain any disconnected subdiagram that has no closed string insertion. In other words, all disconnected components of the worldsheet diagram that have no closed string insertion can be absorbed into the measure factor $\mu$. As always, the overall $e^{-n/g_s}$ due to the instanton action may also be interpreted as the exponentiation of the empty disc diagram. At order $g_s^0$, $\mu$ is computed by the exponentiation of the cylinder diagram,
\ie\label{mufact}
\mu(x_1,\cdots, x_n) = \exp\left[ \sum_{1\leq i, j\leq n} \tikz[baseline={([yshift=-.5ex]current bounding box.center)}]{\pic{cyl2={i}{j}}} +\, {\cal O}(g_s) \right].
\fe
We will see in that in the limit where $x_i$'s are close to one another, $\mu$ approaches the Vandermonde determinant squared $\prod_{i<j} x_{ij}^2$ up to an overall normalization, as expected of the non-Abelian nature of open string modes on coincident D-instantons. The cylinder diagram with the same boundary condition on the two boundary components, namely $i=j$, gives a constant factor that is naively divergent and must be regularized. We will fix this constant factor by matching with the proposed matrix model dual.

An important issue is the choice of integration contour in the $x_i$'s. Conventional view of instantons as Euclidean saddle solutions may suggest integration over Euclidean times for the ZZ-instantons, i.e. purely imaginary $x_i$. However, when a pair of ZZ-instantons of $(1,1)$ type are separated by $2\pi \sqrt{\A'}$ in Euclidean time, the open string ``tachyon" stretched between the two ZZ-instantons becomes on-shell, giving rise to a pole in $\mu(x_1,\cdots, x_n)$. A contour prescription is necessary to render the integration over collective coordinates well defined. Our prescription will be simply to integrate along real $x_i$'s, i.e. Lorentzian times, rather than Euclidean times.

The above prescription generalizes straightforwardly to the case of instantons of type $\{m_1, \cdots, m_\ell\}$, with the appropriately modified integration measure computed from the cylinder diagram. One novelty is that a ZZ-instanton of type $(m, 1)$ comes with a normalization factor ${\cal N}_{m}$ in the integration measure over its coordinate. We do not know of a way to fix ${\cal N}_{m}$ a priori from the worldsheet perspective (say by regularizing the cylinder diagram), and will instead fix them by comparing the answer with the dual matrix model. 
Strikingly, after fixing ${\cal N}_{m}$, we will find precise agreement with the non-perturbative terms in the matrix model result for $1\to k$ amplitude at the 2-instanton level (order $e^{-2/g_s}$), and for $1\to 1$ amplitude at the $n$-instanton level for all $n$. 

The paper is organized as follows. In section \ref{sec:mqm}, we recap the proposal for the non-perturbative completion of the $c=1$ matrix quantum mechanics in \cite{Balthazar:2019rnh}, and explicitly evaluate what would amount to the $n$-instanton contribution to closed string amplitudes. In section \ref{sec:zzinst}, we review the definition of ZZ-instantons, compute the instanton measure from the cylinder diagram, and compute the $n$-instanton contribution to closed string amplitudes for $n=2,3,4$. In section \ref{sec:allorders}, we extend the computation of the closed string reflection amplitude to all instanton orders. We comment on the lessons and implications of our results in section \ref{sec:discussion}.

\section{Instanton expansion in the non-perturbative completion of the $c=1$ matrix quantum mechanics}
\label{sec:mqm}

The closed string sector of $c=1$ string theory, at the perturbative level, has long been conjectured to be dual to a $U(N)$-gauged matrix quantum mechanics in a suitable $N\to \infty$ limit \cite{Klebanov:1991qa,Ginsparg:1993is,Jevicki:1993qn,Martinec:2004td,McGreevy:2003kb}. We refer to the latter as the ``$c=1$ matrix quantum mechanics." It is defined by the Hilbert space of wave functions $\Psi(X)$ in the $N\times N$ Hermitian matrix $X$ that are invariant under the $U(N)$ adjoint action on $X$, with the Hamiltonian $H=\frac{1}{2}\text{Tr}\left(P^2-X^2\right)$, where $P$ is the matrix of canonically conjugate momenta. 
Writing $X=\Omega^{-1}\Lambda\Omega$, where $\Lambda=\text{diag}(\lambda_1,...,\lambda_N)$ and $\Omega\in U(N)$, the wave function $\Psi(X)$ can be expressed as a function of the eigenvalues $\Psi(\{\lambda_i\})$ that is invariant with respect to permutation on the $\lambda_i$'s. The Hamiltonian $H$ acting on the $U(N)$-invariant wave function may be expressed as $H = \Delta^{-1} \hat H \Delta$, where $\Delta= \prod_{i<j}\left(\lambda_i-\lambda_j\right)$, and $\hat H=\frac{1}{2}\sum_{i=1}^N\left(-\partial_{\lambda_i}^2-\lambda_i^2\right)$. The system is thus equivalently described by the wave function $\hat{\Psi}(\{\lambda_i\})\equiv \Delta \Psi(\{\lambda_i\})$ which is completely antisymmetric under permutation of the $\lambda_i$'s, subject to the Hamiltonian $\hat H$. In other words, the system describes $N$ non-relativistic non-interacting fermions in the potential $V(x) = -\frac{1}{2}x^2$.

The appropriate infinite $N$ limit, also known as the double-scaling limit, is defined by taking $N\to\infty$ while keeping the energy of the fermi surface $-\mu$($<0$) finite. In the duality with $c=1$ string theory, $\mu$ is related to the string coupling $g_s$ by $g_s=(2\pi\mu)^{-1}$. In the semiclassical limit $\mu\gg 1$, the closed string vacuum corresponds to the matrix model state in which the fermions fill the right side of the potential, i.e. the region $x>0$, up to the fermi energy $-\mu$. Closed string states are dual to collective excitations of the fermi surface, which may also be viewed as particle-hole pairs. The S-matrix of the collective excitations is most conveniently computed by combining the reflection amplitudes of the individual fermions and holes in the fermi sea \cite{Moore:1991zv}. 

At the non-perturbative level, such a description of the matrix model dual is imprecise. Various proposals of the non-perturbative completion of the matrix model have been considered in the past \cite{Ginsparg:1993is, Takayanagi:2003sm,Douglas:2003up}, either by modifying the Hamiltonian or modifying the notion of the closed string vacuum state. In \cite{Balthazar:2019rnh}, we proposed a specific matrix model state as the dual of the closed string vacuum. This proposal is supported by a detailed agreement of non-perturbative corrections to the closed string scattering amplitudes on both sides of the duality, which we now review.

At any given energy $E$, there are two linearly independent single-fermion eigenstates $\ket{E}_R$ and $\ket{E}_L$ of Hamiltonian $\frac{1}{2}\left(-\partial_{x}^2-x^2\right)$. $\ket{E}_R$ is defined by the fermion wave function with no incoming flux from $x=-\infty$, and $\ket{E}_L$ is the state related by $x\to -x$. The proposed dual of the closed string vacuum state, $|\Omega\rangle$, is one in which the fermions occupy all $\ket{E}_R$ for $E\leq-\mu$ and none other. 

As a scattering state of a non-relativistic fermion, $\ket{E}_R$ has reflection amplitude \cite{Moore:1992gb}
\ie
R(E) = i \mu^{iE} \left[ {1\over 1+e^{2\pi E}}\cdot {\Gamma({1\over 2}-i E)\over \Gamma({1\over 2} + i E)} \right]^{1\over 2}.
\label{exactre}
\fe
The reflection amplitude of a hole (``unoccupation of $\ket{E}_R$ by a fermion") is given by $\left(R(E)\right)^{-1}$. Note that $|R(E)|<1$ due to the tunneling of the fermion through the potential barrier. The same tunneling effect enhances the reflection amplitude of a hole, $\left(R(E)\right)^{-1}$, to have magnitude greater 1. This is in contrast to the type 0B matrix model \cite{Takayanagi:2003sm,Douglas:2003up}, or the theory of ``type II'' in \cite{Ginsparg:1993is}, where both sides of the potential are filled by the fermi sea and the reflection amplitude of a hole is $(R(E))^*$.

The exact $1\to k$ S-matrix element of the closed strings/collective modes, computed using the particle-hole formalism \cite{Moore:1991zv}, takes the form
\ie
S_{1\to k}(\omega,\omega_1,...\omega_k)=\delta\left(\omega-\sum_{i=1}^k\omega_i\right) {\cal{A}}_{1\to k}(\omega_1,...\omega_k),
\fe
with
\ie
{\cal A}_{1\to k}(\omega_1,\cdots, \omega_k)= - \sum_{S_1\sqcup S_2=S}(-1)^{|S_2|}\int_0^{\omega(S_2)}dx \, R(-\mu+\omega-x)(R(-\mu-x))^{-1}.
\label{ammoneton}
\fe
Here $S_1,S_2$ are disjoint subsets of $S=\{\omega_1,...,\omega_k\}$ satisfying $S_1\sqcup S_2=S$, $|S_2|$ denotes the number of elements of $S_2$, and $\omega(S_2)$ is the sum of all elements of $S_2$. The integrand of (\ref{ammoneton}) can be written as
\ie\label{rrfac}
R(-\mu+\omega-x)(R(-\mu-x))^{-1} = \left[ {1+ e^{-2\pi\mu} e^{-2\pi x} \over 1+ e^{-2\pi\mu} e^{2\pi(\omega-x)}} \right]^{1\over 2} K(\mu, \omega, x),
\fe
where
\ie
K(\mu, \omega, x) &\equiv \mu^{i\omega} \left[ {\Gamma({1\over 2} - i (-\mu+\omega-x)) \over \Gamma({1\over 2} - i (-\mu - x))} {\Gamma({1\over 2} + i (-\mu-x))\over \Gamma({1\over 2} + i(-\mu+\omega-x))} \right]^{1\over 2}.
\label{eq:Kfn}
\fe
In (\ref{rrfac}), we have separated a prefactor that captures non-perturbative corrections (suppressed by $e^{-2\pi \mu}$) from the function $K(\mu, \omega, x)$ that captures the Borel resummation of the perturbative asymptotic expansion in $\mu^{-1}$. 
In fact, ${\cal A}_{1\to k}$ admits an instanton expansion of the form 
\ie\label{aonekexp}
{\cal{A}}_{1\to k}(\omega_1,...\omega_k)&= \sum_{g=0}^{\infty}\frac{1}{\mu^{k-1+2g}}{\cal A}^{\text{pert},(g)}_{1\to k}(\omega_1,...\omega_k)\\
&~~~ +\sum_{n=1}^\infty e^{-2\pi n \mu}\sum_{L=0}^\infty \frac{1}{\mu^{{L}}}{\cal A}^{n-\text{inst},(L)}_{1\to k}(\omega_1,...\omega_k),
\fe
where ${\cal A}^{\text{pert},(g)}$ is to be identified with the perturbative closed string amplitude at genus $g$, and ${\cal A}^{n-\text{inst},(L)}$ is the perturbative expansion of the $n$-instanton amplitude at $L$-th {\it open string loop} order. In this case, the perturbative series at each instanton order is Borel-summable, and the summation over $g$ or $L$ in (\ref{aonekexp}) is understood to be the Borel-resummation of the corresponding asymptotic series.



Explicitly, the first few perturbative amplitudes are\footnote{An ambiguity in the definition of asymptotic states of the massless particles in $1+1$d is encountered in the $2\to 2$ amplitude. This ambiguity is resolved by assigning a small imaginary part to $\omega_i$. $\mathrm{Imax}(\{\omega_i\})$ is defined as the $\omega_i$ with largest imaginary part.
The non-analyticity in the $2\to2$ amplitude can be understood as due to the exchange of an on-shell particle \cite{Balthazar:2017mxh}.} 
\ie
&{\cal A}^{\text{pert},(0)}_{1\to 1}(\omega)=\omega,\\
&{\cal A}^{\text{pert},(1)}_{1\to 1}(\omega)=\frac{1}{24}\left(i\omega^2+2i\omega^4-\omega^5\right),\\
&{\cal A}^{\text{pert},(0)}_{1\to 2}(\omega_1,\omega_2)=i \omega\omega_1\omega_2,\\
&{\cal A}^{\text{pert},(0)}_{1\to 3}(\omega_1,\omega_2,\omega_3)=i \omega\omega_1\omega_2\omega_3(1+i\omega),\\
&{\cal A}^{\text{pert},(0)}_{2\to 2}(\{\omega_1,\omega_2\}\to\{\omega_3,\omega_4\})=i \omega_1\omega_2\omega_3\omega_4\left(1+i\mathrm{Imax}(\{\omega_i\})\right),
\fe
where we wrote $\omega\equiv\sum_{i=1}^k \omega_i$ for the total energy in the $1\to k$ amplitude.
These amplitudes were reproduced (numerically, in the case of 4-point tree-level and 2-point one-loop) from the worldsheet formulation of $c=1$ string theory in \cite{Balthazar:2017mxh}.

At one-instanton level, the leading matrix model $1\to k$ amplitude and the subleading $1\to 1$ amplitude are evaluated from (\ref{ammoneton}), (\ref{rrfac}) to be
\ie
{\cal A}^{1-\text{inst},(0)}_{1\to k}&(\omega_1,...,\omega_k)=-\frac{2^{k+1}}{4\pi}\sinh(\pi\omega)\prod_{i=1}^k
\sinh(\pi\omega_i),\\
{\cal A}^{1-\text{inst},(1)}_{1\to 1}&(\omega)= - \frac{i}{2\pi^2}\omega\left(\frac{\pi\omega}{\tanh(\pi\omega)} - 1\right)\sinh^2(\pi\omega).
\fe
In $c=1$ string theory, the non-perturbative corrections to the closed string amplitudes are understood as effects of ZZ-instantons \cite{Zamolodchikov:2001ah, Balthazar:2019rnh}. In the worldsheet formalism, the ZZ-instanton amounts to introducing boundaries to the worldsheet, together with an appropriate integration over the moduli space of the conformal boundary conditions, as will be discussed in section \ref{sec:zzinst}. Such a computation reproduces ${\cal A}^{1-\text{inst},(0)}_{1\to n}$ analytically and ${\cal A}^{1-\text{inst},(1)}_{1\to 1}$ numerically \cite{Balthazar:2019rnh}. This agreement can also be viewed as a non-trivial check of the proposed matrix model state dual to the closed string vacuum.

The leading non-perturbative contribution to the $1\to k$ amplitude (\ref{ammoneton}) at the $n$-instanton level is given by
\ie\label{aonegenform}
&{\cal A}^{n-\text{inst},(0)}_{1\to k}(\omega_1,...,\omega_k)
\\
&=\frac{1}{2\pi^\frac{3}{2}}\frac{(-1)^n}{n}\frac{\Gamma\left(\frac{1}{2}+n\right)}{\Gamma\left(1+n\right)}e^{\pi\omega n} {}_{2}F_{1}\left(-1/2,-n,1/2-n,e^{-2\pi\omega}\right)  2^k\prod_{i=1}^k\sinh(n\pi  \omega_i).
\fe
The main objective of this paper is to understand how (\ref{aonegenform}) arises from the worldsheet perspective for $n\geq 2$. In section \ref{sec:zzinst}, we will extend the formalism of \cite{Balthazar:2019rnh} to calculate the effect of multiple ZZ-instantons on the closed string scattering amplitudes, and reproduce the following special cases:
\ie{}
&{\cal A}^{2-\text{inst},(0)}_{1\to k}(\omega_1,\cdots,\omega_k)=\frac{2^{k}}{8\pi}\sinh(\pi\omega)\left[2\cosh(\pi\omega)+\sinh(\pi\omega)\right]\prod_{i=1}^k\sinh(2\pi\omega_i),\\
& {\cal A}^{3-\text{inst},(0)}_{1\to 1}(\omega)=-\frac{1}{12\pi}\sinh^2(\pi\omega)\left[4+5\cosh(2\pi\omega)+3\cosh(4\pi\omega)+2\sinh(2\pi\omega)+2\sinh(4\pi\omega)\right],\\
& {\cal A}^{4-\text{inst},(0)}_{1\to 1}(\omega )= \frac{1}{128\pi}\sinh^2(\pi\omega)\left[32+44\cosh(2\pi\omega)+32\cosh(4\pi\omega)\right.\\
&~~~~~~~~~~~~~~~~~~~~~~~\left.+20\cosh(6\pi\omega)+15\sinh(2\pi\omega)+18\sinh(4\pi\omega)+15\sinh(6\pi\omega)\right].
\label{eq:1to1MM}
\fe
A number of combinatorial observations based on these computations will then allow us to derive, in section \ref{sec:allorders}, ${\cal A}_{1\to 1}^{n-{\rm inst},(0)}$ for all $n$ from the ZZ-instanton computation.

\section{Multiple ZZ-instantons from the worldsheet perspective}
\label{sec:zzinst}

The worldsheet formulation of $c=1$ string theory is based on the CFT that consists of $c=25$ Liouville theory, a timelike free boson $X^0$, and the $bc$ conformal ghost system. The Virasoro primaries of the $c=25$ Liouville CFT are scalar operators $V_{P}$ labeled by their Liouville momenta $P\geq 0$, of scaling dimensions $\Delta_P=2+2P^2$, subject to the normalization
\ie
\langle V_{P_1}(z,\bar z)V_{P_2}(0)\rangle=\pi\frac{\delta(P_1-P_2)}{|z|^{2\Delta_{P_1}}}
\fe
Their structure constants are given by the DOZZ formula \cite{Dorn:1994xn,Zamolodchikov:1995aa}.\footnote{Explicitly, in our normalization convention, the $c=25$ Liouville structure constants are given in (2.9) of \cite{Balthazar:2017mxh}.} 
The closed string asymptotic states, as insertions on the worldsheet, are given by the BRST cohomology representatives of the form
\ie
{\cal V}^{\pm}_\omega= g_s c\tilde{c} \, e^{\pm i \omega X^0} V_{P=\frac{\omega}{2}},
\fe
where the superscript $+$ and $-$ corresponds to in- and out-states, respectively. Sometimes referred to as ``tachyons" in the literature for historical reasons, these closed string excitations behave as 1+1 dimensional massless particles in the asymptotic (weak coupling) region of spacetime.

The perturbative closed string amplitudes are computed by integrating appropriate correlation functions of vertex operators and $b$-ghost insertions over the moduli space of punctured Riemann surfaces, as in the usual bosonic string theory. Unlike the critical bosonic string theory which suffers from closed string tachyon divergence at loop levels, the perturbative $c=1$ string amplitudes are perfectly finite and compatible with perturbative unitarity \cite{Balthazar:2017mxh}. Assuming the duality with the matrix model, which is checked up to 1-loop order in \cite{Balthazar:2017mxh}, the perturbative series of the $c=1$ string amplitude is in fact Borel summable. However, the Borel-resummed perturbative amplitude by itself does not admit the interpretation as the scattering amplitude of collective excitations of free non-relativistic fermions \cite{Balthazar:2019rnh}.

Following the general prescription of \cite{Polchinski:1994fq}, one expects non-perturbative corrections to the closed string amplitude due to D-instantons. Namely, one considers worldsheets with boundaries, subject to conformal boundary conditions that describe strings ending on D-instantons, and integrate over the moduli space of Riemann surface with boundaries, as well as over the moduli space of boundary conditions i.e. the D-instanton moduli space. The one-instanton contribution to closed string scattering was studied in type IIB string theory in \cite{Green:1997tv}, and in $c=1$ string theory in \cite{Balthazar:2019rnh}. In the latter case, the relevant D-instantons are described by ZZ-boundary condition in Liouville theory \cite{Zamolodchikov:2001ah} and Dirichlet boundary condition in $X^0$. Delicate cancelations between worldsheet diagrams of different topologies are seen to render the instanton amplitudes well defined and agree with the proposed matrix model dual \cite{Balthazar:2019rnh, Sen:2019qqg}. Multi-instanton contributions, as will be discussed below, are subject to further complications in the integration over the instanton moduli space.

Firstly, as outlined in the introduction, at the $n$-instanton level ($n\geq 2$), there can be different types of ZZ-instantons that do not lie in a single connected moduli space of exactly marginal deformations of boundary conditions, all of which contribute to the order $e^{-n/g_s}$ closed string amplitude. These will be described in detail in section \ref{sec:c1bcs}. Secondly, there is a nontrivial integration measure factor over the instanton moduli space, computed by the vacuum diagram with boundaries ending on the ZZ-instantons. We will see that the measure factor develops a pole when an open string mode stretched between a pair of ZZ-instanton becomes on-shell (or ``massless"). A contour prescription for handling the integration near the poles will be given in section \ref{sec:cyl}. We will apply this prescription to compute the closed string amplitudes at $n=2, 3, 4$ instanton levels and find remarkable agreement with the matrix model result (\ref{eq:1to1MM}).


\subsection{ZZ boundary conditions and instantons}
\label{sec:c1bcs}

Conformal boundary conditions of Liouville CFT come in two types: FZZT \cite{Fateev:2000ik, Teschner:2000md} and ZZ \cite{Zamolodchikov:2001ah}. The former corresponds to a semi-infinite partially-space-filling brane, whereas the latter corresponds a point-like brane localized in the strong coupling region. In this work we are concerned with D-instantons of finite action, described by ZZ boundary condition in the Liouville CFT tensored with Dirichlet boundary condition in $X^0$, and direct sums thereof. 

It was shown in \cite{Zamolodchikov:2001ah} that there is a discrete family of ZZ boundary conditions, which we refer to as the $(m,n)$-type ZZ boundary condition, labeled by a pair of positive integers $m$ and $n$. In a unitary Liouville theory, only the $(1,1)$ ZZ boundary condition supports a unitary spectrum of boundary operators. This gives rise to the so-called ZZ-branes which have been discussed extensively in the context of $c=1$ string theory \cite{McGreevy:2003kb,Klebanov:2003km}. The ZZ-instanton constructed from the $(1,1)$ ZZ boundary condition was considered in the one-instanton analysis of \cite{Balthazar:2019rnh}. At the multi-instanton level, we will see that the $(m, n)$ ZZ boundary conditions give rise to a more general class of ZZ-instantons whose effect on closed string amplitudes should be taken into account. 

The $(1,1)$ ZZ boundary condition may be defined as the conformal boundary condition in Liouville CFT that supports the identity operator as the only boundary Virasoro primary. The $(m,n)$ ZZ boundary condition has the property that the only boundary Virasoro primary that interpolates between the $(1,1)$ and $(m,n)$ ZZ boundary condition corresponds to the degenerate representation of the boundary Virasoro algebra labeled by $(m,n)$. In the $c=25$ case, such a degenerate primary has weight $1-{(m+n)^2\over 4}$, and Virasoro character
\ie
\chi_{(m,n)}(\tau)=\frac{q^{-\frac{(m+n)^2}{4}}-q^{-\frac{(m-n)^2}{4}}}{\eta(\tau)},
\label{char}
\fe
where $\eta(\tau)$ is the Dedekind eta-function and $q=e^{2\pi i \tau}$. The $(m,n)$-type ZZ boundary state takes the form
\ie
|{\rm ZZ} (m,n)\rangle\rangle_{\rm Liouville} =\int_0^\infty\frac{dP}{\pi}\Psi^{(m,n)}(P)|V_P\rangle\rangle,
\fe
where $|V_P\rangle\rangle$ is the Ishibashi state constructed from the bulk Liouville primary $V_P$. Consideration of the cylinder partition function with (1,1) ZZ boundary condition on one side and $(m,n)$ on the other,
\ie
\chi_{(m,n)}\left( {\tau}\right)=\int_0^\infty\frac{dP}{\pi}\Psi^{(m,n)}(P)\Psi^{(1,1)}(P)\chi_{1+P^2}\left(-1/\tau\right)
\label{eq:mncyl}
\fe
where $\chi_h(\tau)$ is the $c=25$ Virasoro character for a primary operator of weight $h$, determines $\Psi^{(m,n)}(P)$ to be
\ie
\Psi^{(m,n)}(P)=2^{\frac{5}{4}}\sqrt{\pi}\frac{\sinh(2\pi m P)\sinh(2\pi n P)}{\sinh(2\pi P)}.
\label{eq:Psimn}
\fe
As (\ref{eq:Psimn}) is invariant under the exchange of $m$ with $n$, we will restrict to $m\geq n$ from now on.

All boundary structure constants can be bootstrapped from crossing relations among boundary correlators \cite{Zamolodchikov:2001ah, Giombi:2008sr}, although we will not make explicit use of them in this paper. The spectrum of boundary operators interpolating between the $(m,n)$ and $(m',n')$ ZZ boundary conditions is given by the cylinder partition function
\ie{}
&\int_0^\infty\frac{dP}{\pi}\Psi^{(m,n)}(P)\Psi^{(m',n')}(P)\chi_{1+P^2}\left(-1/\tau\right)
\\
&~~~ =\sum_{p=0}^{{\rm{min}} (m,m')-1}\sum_{q=0}^{{\rm{min}} (n,n')-1} \chi_{(m+m'-2p-1,n+n'-2l-1)}\left({\tau}\right).
\label{eq:Zmn}
\fe

In $c=1$ string theory, a single ZZ-instanton of type $(m,n)$ located at time $X^0 = x$ is described by the matter CFT boundary state $|{\rm ZZ} (m,n)\rangle\rangle_{\rm Liouville}\otimes |D(x) \rangle\rangle_{X^0}$. More generally, one can consider direct sums of such boundary states. The action of the $(1,1)$ ZZ-instanton, $S_{(1,1)}$, is related to the mass of the $(1,1)$ ZZ-brane $M_{(1,1)}$ by \cite{McGreevy:2003kb, Balthazar:2019rnh}
\ie
S_{(1,1)}=2\pi M_{(1,1)}=\frac{1}{g_s}.
\label{S11}
\fe
Upon analytic continuation to $P\to i$ (so that $\Delta_P\to 0$), the disc 1-point function $\Psi(P) = \langle V_P| ZZ\rangle\rangle$ is proportional to the ``empty disc" diagram which can be identified with minus the instanton action. This allows us to determine the action of the $(m,n)$ ZZ-instanton to be 
\ie
S_{(m,n)}=\lim_{P\to i}\frac{\Psi^{(m,n)}(P)}{\Psi^{(1,1)}(P)} S_{(1,1)}=\frac{m  n }{g_s}.
\label{eq:Smn}
\fe
For later use we also record the disc 1-point diagram with the closed string insertion ${\cal{V}}^\pm_{\omega}$, with boundary on the $(m,n)$ ZZ-instanton (generalizing (2.8) of \cite{Balthazar:2019rnh}),
\ie
\tikz[baseline={([yshift=-1.6ex]current bounding box.center)}]{\pic{diskbig={\{(m,n),x\}}{\omega}}}=\langle {\cal{V}}^\pm_{\omega}\rangle^{D_2}_{(m,n),x}=g_s\frac{C_{D_2}}{2\pi}\Psi^{(m,n)}\left(\frac{|\omega|}{2}\right)e^{\pm i\omega x}=2e^{\pm i\omega x} \frac{\sinh(m \pi\omega) \sinh(n \pi\omega)}{\sinh(\pi\omega)}.
\label{eq:disk1pt}
\fe

In fact, we will see in section \ref{sec:4zz} that, agreement with the matrix model dual suggests that only the $(m,1)$ ZZ-instantons give rise to non-perturbative corrections to the closed string amplitudes of consideration. A multi-ZZ-instanton configuration described by the direct sum of boundary states $|{\rm ZZ} (m_i,1)\rangle\rangle_{\rm Liouville}\otimes |D(x_i) \rangle\rangle_{X^0}$, $i=1,\cdots, \ell$, will be referred to as an instanton of type $\{m_1,...,m_\ell\}$. The moduli space of such instantons is parameterized by the collective coordinates $x_1,\cdots, x_\ell$.

\subsection{The instanton measure}
\label{sec:cyl}

The instanton-mediated non-perturbative correction to the closed string amplitude is computed by worldsheet diagrams with boundaries on the D-instantons, integrated over the instanton moduli space in the form (\ref{schminst}), with a suitable measure factor $\mu$ that is a function of the instanton collective coordinates. Unlike in the path integral formulation of quantum field theories, where the instanton measure can be derived by integrating over fluctuations around the instanton solution, such a derivation is not available for the D-instanton. Nonetheless, one expects that the instanton measure $\mu$ is computed by exponentiating open string vacuum diagrams of one-loop and higher orders, as (\ref{mufact}). In the one-instanton case considered in \cite{Balthazar:2019rnh}, the measure factor is a constant (by time-translation invariance), and may be viewed as a renormalization of the instanton action. In the multi-instanton case, however, the measure factor $\mu$ depends nontrivially on the relative position of the ZZ-instantons in Euclidean time. At order $g_s^0$, $\mu$ is computed by exponentiating the cylinder diagram, which we will now analyze.


One class of cylinder diagrams has both boundaries on the same ZZ-instanton, say of type $(m,n)$. Such diagrams are formally independent of the instanton collective coordinate $x$, and is furthermore divergent. We do not know of a canonical regularization scheme of such diagrams in the worldsheet formalism. Instead, we will assume that such diagrams can be absorbed into an overall normalization constant ${\cal N}_{(m,n)}$ associated with the integration over the collective coordinate of the ZZ-instanton of type $(m,n)$, and will determine ${\cal N}_{(m,n)}$ by comparison with the dual matrix model.\footnote{
We will see that ${\cal N}_{(m,n)}$ are real normalization factors associated with the integration in Lorentzian collective coordinates of the ZZ-instantons. This is in contrast to the Euclidean partition function of the ZZ-instanton which receives an imaginary contribution due to the open string tachyon.
} In fact, we will find that only ZZ-instantons of type $(m,1)$ contribute, and will use the notation ${\cal N}_{(m,1)}\equiv {\cal N}_m$.

The cylinder diagrams with two boundaries on different ZZ-instantons (which may or may not be of the same type), on the other hand, can be evaluated unambiguously. Let us begin by considering the cylinder diagram between two $(1,1)$ ZZ-instantons located at Euclidean times $x_1^E$ and $x_2^E$ respectively. The free boson cylinder partition function is given by $e^{- t\frac{(\Delta x^E)^2}{2\pi}}/\eta(i t)$, where $t$ parameterizes the modulus of the cylinder and $\Delta x^E \equiv x_1^E-x_2^E$ is the separation of the two ZZ-instantons in Euclidean time. The Liouville cylinder partition function with ZZ boundary condition is $(e^{2\pi t}-1)/\eta(it)$, as follows from a special case of (\ref{char}) and (\ref{eq:Zmn}). Combining with the $bc$ ghost contribution $\eta(it)^2$, we obtain the cylinder amplitude 
\ie
 \tikz[baseline={([yshift=-1.5ex]current bounding box.center)}]{\pic{cyl11={1}{2}}}
 =\int_0^\infty \frac{dt}{2t}\left(e^{2\pi  t}-1\right)e^{-t\frac{(\Delta x^E)^2}{2\pi}}=\frac{1}{2}\ln\left(\frac{(\Delta x^E)^2}{(\Delta x^E)^2-4\pi^2}\right),
\label{cyl11}
\fe
where the moduli integral is performed with the assumption $|\Delta x^E|>2\pi$. When $|\Delta x^E|<2\pi$, the lowest open string mode stretched between the two ZZ-instantons becomes ``tachyonic", and we will define the cylinder amplitude by analytic continuation from the $|\Delta x^E|>2\pi$ regime. Exponentiating the cylinder amplitude as in (\ref{mufact}) then gives the order $g_s^0$ measure factor on the moduli space of two $(1,1)$ ZZ-instantons,
\ie
\mu(x_1, x_2) = {\cal N}_1^2\frac{(\Delta x^E)^2}{(\Delta x^E)^2-4\pi^2} .
\label{meas}
\fe
Furthermore, a symmetry factor ${1\over 2}$ should be included due to the indistinguishability of the two ZZ-instantons. 
Note that in the $\Delta x^E \to 0$ limit, the factor $(\Delta x^E)^2$ in the measure can be interpreted as the Vandermonde determinant in gauge fixing the non-Abelian coordinate of two instantons to the diagonal form.

At $\Delta x^E = \pm 2\pi$, the stretched open string mode becomes on-shell and (\ref{meas}) develops a pole. A contour prescription is needed to define the eventual integration over instanton collective coordinates in a way that circumvents the pole. Our prescription will be simply to analytic continue (\ref{meas}) to Lorentzian $\Delta x$, and integrate the worldsheet diagram along the real $x_i$-contour.\footnote{An alternative contour prescription that integrates in Euclidean times while circumventing the poles appears to give the same results, with a different assignment of the normalization constants ${\cal N}_{(m,1)}$ for the $(m,1)$ ZZ-instantons, $m\geq 2$.}



A similar analysis extends to pairs of ZZ-instantons of the more general type $(m,n)$. For example, the cylinder diagram between a (1,1) and an $(n,1)$ ZZ-instanton (with $n\neq1$) evaluates to
\ie
\tikz[baseline={([yshift=-1.5ex]current bounding box.center)}]{\pic{cyln1={1}{2}}}=\frac{1}{2}\int_0^\infty \frac{dt}{t}\left[ e^{2\pi  t (n+1)^2/4}-e^{2\pi  t (n-1)^2/4}\right] e^{-t\frac{(\Delta x^E)^2}{2\pi}}=\frac{1}{2}\ln\left[\frac{(\Delta x^E)^2 - ((n-1)\pi)^2}{(\Delta x^E)^2 - ((n+1)\pi)^2}\right],
\label{cyl11andn1}
\fe
where the dashed and solid boundaries on the LHS correspond to the $(n,1)$ and $(1,1)$ ZZ boundary conditions respectively. On the other hand, the cylinder diagram between a pair of $(n,1)$ ZZ-instantons gives
\ie
\tikz[baseline={([yshift=-1.5ex]current bounding box.center)}]{\pic{cylnn={1}{2}}}=\frac{1}{2}\int_0^\infty \frac{dt}{t}\left(e^{2\pi  t n^2}-1\right)e^{-t\frac{(\Delta x^E)^2}{2\pi}}=\frac{1}{2}\ln\left[\frac{(\Delta x^E)^2}{(\Delta x^E)^2 - (2n\pi)^2}\right].
\label{cyln1andn1}
\fe
With a Lorentzian integration contour in the $x_i$'s, all poles in the measure factor are avoided.

\subsection{2-instanton corrections to the closed string $1\to k$ amplitude}
\label{sec:2zz}

We begin by considering the leading correction to the closed string $1\to 1$ (reflection) amplitude at the 2-ZZ-instanton level, of order $e^{-2/g_s}$. There are two types of contributions, namely two $(1,1)$ ZZ-instantons, and a single $(2,1)$ ZZ-instanton.

In the case of two $(1,1)$ ZZ-instantons, the worldsheet diagram consists of two discs, each with one closed string vertex operator insertion. The boundaries of the two discs may lie on the two separate ZZ-instantons at times $x_1$ and $x_2$, or both boundaries may lie on the same ZZ-instanton, either at $x_1$ or at $x_2$. The relevant disc 1-point diagram is evaluated in (\ref{eq:disk1pt}). We then integrate over $x_1$, $x_2$ with the measure factor (\ref{cyl11andn1}) along the Lorentzian contour. 

The contribution from the two discs ending on the same ZZ-instanton, say the one at $x_1$, is given by
\ie\label{ampsddo}
&e^{-\frac{2}{g_s}}\cN_1^2 \frac{1}{2} 2^2 \sinh(\pi\omega_1)\sinh(\pi\omega_2)\int dx_1 dx_2 \frac{\Delta x^2}{\Delta x^2 + (2\pi)^2} e^{i (\omega_1-\omega_2)x_1}  \\
&= e^{-\frac{2}{g_s}}\cN_1^2  4\pi \delta(\omega_1 - \omega_2) \sinh^2(\pi\omega_1)\int d\Delta x \frac{\Delta x^2}{\Delta x^2 + (2\pi)^2}.
\fe
However, the integration over large $\Delta x$ is linearly divergent. This is in fact due to an over-counting. Namely, we should normalize all amplitudes by the vacuum amplitude, which itself contains ZZ-instanton contributions. We must then subtract from (\ref{ampsddo}) a ``disconnected two-instanton amplitude" in which the second ZZ-instanton merely contributes to the vacuum amplitude. This amounts to replacing the integrand on the RHS of (\ref{ampsddo}) by
\ie
\frac{\Delta x^2}{\Delta x^2 + (2\pi)^2} \to \frac{\Delta x^2}{\Delta x^2 + (2\pi)^2} -1.
\fe
There is an identical contribution coming from both discs ending on the ZZ-instanton at $x_2$.

The contribution from two discs ending on the two separate ZZ-instantons, on the other hand, is given by
\ie{}
&e^{-\frac{2}{g_s}}\cN_1^2 \frac{1}{2} 2^2 \sinh(\pi\omega_1)\sinh(\pi\omega_2)\int dx_1 dx_2  \frac{\Delta x^2}{\Delta x^2 + (2\pi)^2} \left( e^{i \omega_1 x_1 - i\omega_2 x_2} + e^{i \omega_1 x_2 - i\omega_2 x_1}\right) \\
&=e^{-\frac{2}{g_s}}\cN_1^2  4\pi \delta(\omega_1 - \omega_2) \sinh^2(\pi\omega_1)\int d\Delta x \frac{\Delta x^2}{\Delta x^2 + (2\pi)^2} \left( e^{i \omega_1 \Delta x} + e^{-i \omega_1 \Delta x} \right).
\fe

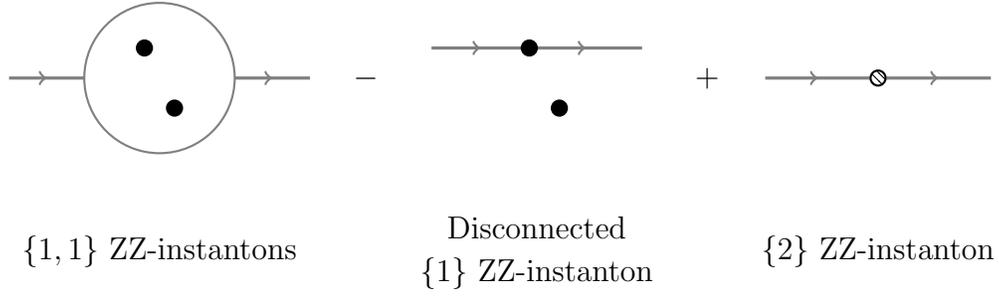
\begin{figure}[h!]
\centering
\begin{tabular}{c c c c c}
\begin{tikzpicture}
[yscale=1,line width=.8pt,
    decoration={
      markings,
      mark=at position 0.5 with {\arrow[line width=1pt]{>}}
    }
]
\draw[color=gray, thick] (0,0) circle [radius=1];
\draw[color=gray,very thick,postaction=decorate] (-2,0) -- (-1,0);
\draw[color=gray,very thick,postaction=decorate] (1,0) -- (2,0);
\draw[fill] (-0.2,0.4) circle [radius=0.1];
\draw[fill] (0.2,-0.4) circle [radius=0.1];
\end{tikzpicture} &
\begin{tikzpicture}
\draw[color=white] (0,-1)--(0,1);
\node at (0,0) {$-$};
\end{tikzpicture} &
\begin{tikzpicture}
[yscale=1,line width=.8pt,
    decoration={
      markings,
      mark=at position 0.5 with {\arrow[line width=1pt]{>}}
    }
]
\draw[color=white, thick] (0,0) circle [radius=1];
\draw[color=gray,very thick,postaction=decorate] (-1.5,0.4) -- (-0.2,0.4);
\draw[color=gray,very thick,postaction=decorate] (-0.2,0.4) -- (-0.2+1.5,0.4);
\draw[fill] (-0.2,0.4) circle [radius=0.1];
\draw[fill] (0.2,-0.4) circle [radius=0.1];
\end{tikzpicture} &
\begin{tikzpicture}
\draw[color=white] (0,-1)--(0,1);
\node at (0,0) {$+$};
\end{tikzpicture} &
\begin{tikzpicture}
[yscale=1,line width=.8pt,
    decoration={
      markings,
      mark=at position 0.5 with {\arrow[line width=1pt]{>}}
    }
]
\draw[color=white, thick] (0,0) circle [radius=1];
\draw[color=gray,very thick,postaction=decorate] (-1.5,0) -- (-0.1,0);
\draw[color=gray,very thick,postaction=decorate] (0.1,0) -- (1.5,0);
\draw[color=white] (0,0) circle [radius=0.1];
\draw[pattern=north west lines, pattern color=black] (0,0) circle [radius=0.1];
\end{tikzpicture} \\
\\
\makecell{$\{1,1\}$ ZZ-instantons} & & 
\makecell{Disconnected\\$\{1\}$ ZZ-instanton}
 & & 
\makecell{$\{2\}$ ZZ-instanton}
\end{tabular}
\caption{Summary of 2-instanton contributions to the $1\to 1$ closed string amplitude. The first diagram represents worldsheets with boundary on the direct sum of two $(1,1)$ ZZ-instantons. The second diagram represents subtraction of the disconnected instanton amplitude in which one of the instantons contributes to the vacuum amplitude. The third diagram represents the $(2,1)$ ZZ-instanton contribution.}
\label{fig:11disc}
\end{figure}

The contribution from a single $(2,1)$ ZZ-instanton comes from two disconnected discs, each with one closed string insertion, subject to the same boundary condition. After integrating out the collective coordinate, the result is
\ie
e^{-{2\over g_s}} \cN_2  8\pi \delta(\omega_1 - \omega_2) \sinh^2(2\pi\omega_1).
\fe
Putting these together, we obtain the total 2-instanton contribution to the closed string $1\to 1$ amplitude,
\ie{}
& e^{-{2\over g_s}} \cN_1^2\frac{1}{2}  \int dx_1 dx_2\, \left[ 2 ~ \tikz[baseline={([yshift=-1.6ex]current bounding box.center)}]{\pic{disk1={1}}}~\tikz[baseline={([yshift=-1.6ex]current bounding box.center)}]{\pic{disk1={1}}} \left( \exp\Big(2~\tikz[baseline={([yshift=-1.6ex]current bounding box.center)}]{\pic{cyl11={1}{2}}}\Big) - 1 \right) + \biggl( \tikz[baseline={([yshift=-1.6ex]current bounding box.center)}]{\pic{disk1={1}}}~\tikz[baseline={([yshift=-1.6ex]current bounding box.center)}]{\pic{disk1={2}}} \,+\, \tikz[baseline={([yshift=-1.6ex]current bounding box.center)}]{\pic{disk1={2}}}~\tikz[baseline={([yshift=-1.6ex]current bounding box.center)}]{\pic{disk1={1}}} \biggr)\exp\Big(2~\tikz[baseline={([yshift=-1.6ex]current bounding box.center)}]{\pic{cyl11={1}{2}}}\Big) \right]\\
& + e^{-{2\over g_s}}\cN_2 \int dx_1 ~  \tikz[baseline={([yshift=-1.6ex]current bounding box.center)}]{\pic{diskn={1}}}~\tikz[baseline={([yshift=-1.6ex]current bounding box.center)}]{\pic{diskn={1}}}\\
&  =  \vphantom{\int} e^{-{2\over g_s}} 2\pi \delta(\omega_1 -\omega_2) \left\lbrace \cN_1^2 2 \sinh^2(\pi\omega_1) \left[ 2\int d\Delta x\left( \frac{(\Delta x)^2}{(\Delta x)^2+(2\pi)^2} - 1 \right) \right. \right.
\\
&~~~~~~\left.\left.+ \int d\Delta x\, \frac{(\Delta x)^2}{(\Delta x)^2+(2\pi)^2}\left( e^{i\omega_1 \Delta x} + e^{-i\omega_1 \Delta x} \right) \right] \right.+ \left.\vphantom{\int} \cN_2 4 \sinh^2(2\pi\omega_1)\right\rbrace \\
&  =  e^{-{2\over g_s}} 2\pi \delta(\omega_1 -\omega_2) \Big[ - \cN_1^2 8\pi^2 \sinh^2(\pi\omega_1) \left( 1+ e^{-2\pi \omega_1} \right) + \cN_2 4\sinh^2(2\pi\omega_1)\Big].
\label{eq:2zz1to1}
\fe
Using $\cN_1=-\frac{1}{8\pi^2}$ \cite{Balthazar:2019rnh}, comparison with the matrix model result (\ref{eq:1to1MM}) yields
\ie
\cN_2 = {3\over 64\pi^2}.
\label{N21}
\fe
It is useful to organize the 2-instanton computation according to Figure \ref{fig:11disc}, where the subtraction of disconnected instanton diagram is indicated. While the subtraction scheme is fairly simple in the 2-instanton case, it will become progressively more complicated at higher instanton numbers.

The generalization of the above computation to $1\to k$ closed string amplitude at order $e^{-2/g_s}$ is straightforward. Let $\omega$ label the total energy, and $\omega_1, \cdots, \omega_k$ the energies of outgoing closed strings. The subtraction of disconnected diagrams is similar to the $1\to1$ case. The contribution from a pair of $(1,1)$ ZZ-instantons to the $1\to k$ amplitude is
\ie{}
& e^{-{2\over g_s}} \cN_1^2\frac{1}{2}  \int dx_1 dx_2\, \left[ \left( \tikz[baseline={([yshift=-1.6ex]current bounding box.center)}]{\pic{disk1w={1}{\omega}}}+\tikz[baseline={([yshift=-1.6ex]current bounding box.center)}]{\pic{disk1w={2}{\omega}}}\right)\times...\times\left( \tikz[baseline={([yshift=-1.6ex]current bounding box.center)}]{\pic{disk1w={1}{\omega_k}}}+\tikz[baseline={([yshift=-1.6ex]current bounding box.center)}]{\pic{disk1w={2}{\omega_k}}}\right) \exp \Big(2~\tikz[baseline={([yshift=-1.6ex]current bounding box.center)}]{\pic{cyl11={1}{2}}}\Big) -2 ~ \tikz[baseline={([yshift=-1.6ex]current bounding box.center)}]{\pic{disk1w={1}{\omega}}} \times...\times\tikz[baseline={([yshift=-1.6ex]current bounding box.center)}]{\pic{disk1w={1}{\omega_k}}} \right]\\
&=e^{-{2\over g_s}} 2\pi \delta\left(\omega-\sum_{i=1}^k\omega_k\right) \cN_1^2 2^{k}\sinh(\pi \omega)\prod_{i=1}^k\sinh(\pi \omega_i)
\\
&~~~~~~\times\int d\Delta x \left[\frac{(\Delta x)^2}{(\Delta x)^2+4\pi^2} \left( 1+ e^{-i \omega \Delta x}\right)\prod_{i=1}^k \left( 1+ e^{i \omega_i \Delta x}\right) -2\right].
\label{eq:2zz1tok}
\fe
The integral in the last line can be evaluated as 
\ie{}
& \int d\Delta x \left[\frac{(\Delta x)^2}{(\Delta x)^2+4\pi^2} \sum_{S_1,S_2}^{S_1\sqcup S_2=S}2\cos {\left(\omega+\omega(S_1)-\omega(S_2) \right) \Delta x\over 2}  ~ -2\right]
\\
&= -4\pi^2 \sum_{S_1,S_2}^{S_1\sqcup S_2=S} e^{-\pi \left(\omega+\omega(S_1)-\omega(S_2) \right) }
= -4\pi^2 e^{-\pi \omega} \prod_{i=1}^k 2\cosh(\pi\omega_i),
\fe
where we have defined $S = \{\omega_1,...,\omega_n\}$, $S_1$, $S_2$ are disjoint subsets of $S$ such that $S_1\sqcup S_2=S$, and $\omega(S_i)=\sum_{\omega_\ell\in S_i}\omega_\ell$. 

The $(2,1)$ ZZ-instanton contribution to the $1\to k$ amplitude evaluates to
\ie
e^{-\frac{2}{g_s}}2\pi\delta\left(\omega-\sum_{i=1}^k\omega_i\right) {\cal N}_2 2^{k+1}\sinh(2\pi\omega)\prod_{i=1}^k\sinh(2\pi\omega_i).
\label{twoinszonek}
\fe
Putting (\ref{eq:2zz1tok}) and (\ref{twoinszonek}) together, and using (\ref{N21}), we recover precisely the matrix model result for ${\cal A}_{1\to k}^{2-{\rm inst},(0)}(\omega_1, \cdots, \omega_k)$ given in (\ref{eq:1to1MM}).

\subsection{3-instanton corrections to the closed string $1\to 1$ amplitude}
\label{sec:3zz}

At order $e^{-3/g_s}$ there are several types of contributions: three $(1,1)$ ZZ-instantons, a $(1,1)$ together with a $(2,1)$ ZZ-instanton, and a single $(3,1)$ ZZ-instanton. We will label these instanton configurations by $\{1,1,1\}$, $\{2,1\}$, and $\{3\}$, respectively. In the following subsections we will evaluate each contribution separately. The relevant worldsheet correlators, namely the disc 1-point function and the cylinder partition function, have been explicitly evaluated in (\ref{eq:disk1pt}) and (\ref{cyl11andn1}).

\subsubsection{$\{1,1,1\}$ ZZ-instantons}

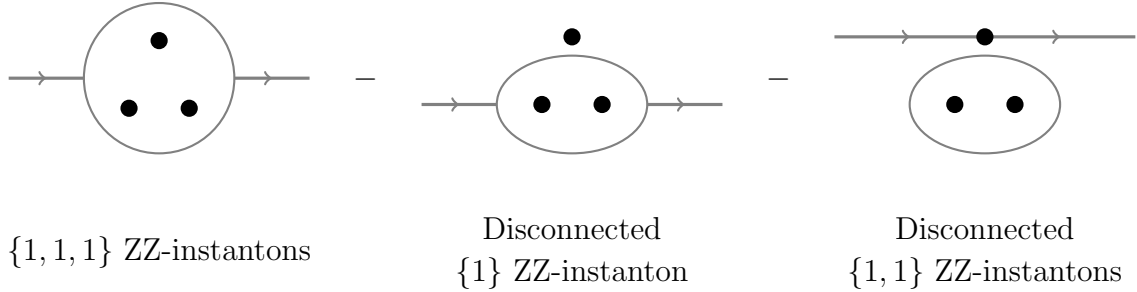
\begin{figure}[h!]
\centering
\begin{tabular}{c c c c c}
\begin{tikzpicture}
[yscale=1,line width=.8pt,
    decoration={
      markings,
      mark=at position 0.5 with {\arrow[line width=1pt]{>}}
    }
]
\draw[color=gray, thick] (0,0) circle [radius=1];
\draw[color=gray,very thick,postaction=decorate] (-2,0) -- (-1,0);
\draw[color=gray,very thick,postaction=decorate] (1,0) -- (2,0);
\draw[fill] (0,0.5) circle [radius=0.1];
\draw[fill] (-0.4,-0.40) circle [radius=0.1];
\draw[fill] (0.4,-0.4) circle [radius=0.1];
\end{tikzpicture} &
\begin{tikzpicture}
\draw[color=white] (0,-1)--(0,1);
\node at (0,0) {$-$};
\end{tikzpicture} &
\begin{tikzpicture}
[yscale=1,line width=.8pt,
    decoration={
      markings,
      mark=at position 0.5 with {\arrow[line width=1pt]{>}}
    }
]
\draw[fill] (0,0.5) circle [radius=0.1];
\draw[fill] (-0.4,-0.40) circle [radius=0.1];
\draw[fill] (0.4,-0.4) circle [radius=0.1];
\draw[color=gray] (0,-0.4) ellipse (1 and 0.65);
\draw[color=gray,very thick,postaction=decorate] (-2,-0.4) -- (-1,-0.4);
\draw[color=gray,very thick,postaction=decorate] (1,-0.4) -- (2,-0.4);
\end{tikzpicture} &
\begin{tikzpicture}
\draw[color=white] (0,-1)--(0,1);
\node at (0,0) {$-$};
\end{tikzpicture} &
\begin{tikzpicture}
[yscale=1,line width=.8pt,
    decoration={
      markings,
      mark=at position 0.5 with {\arrow[line width=1pt]{>}}
    }
]
\draw[color=gray] (0,-0.4) ellipse (1 and 0.65);
\draw[color=gray,very thick,postaction=decorate] (-2,0.5) -- (0,0.5);
\draw[color=gray,very thick,postaction=decorate] (0,0.5) -- (2,0.5);
\draw[fill] (0,0.5) circle [radius=0.1];
\draw[fill] (-0.4,-0.40) circle [radius=0.1];
\draw[fill] (0.4,-0.4) circle [radius=0.1];
\end{tikzpicture} \\
\\
\makecell{$\{1,1,1\}$ ZZ-instantons} & & 
\makecell{Disconnected\\$\{ 1 \}$ ZZ-instanton} & & 
\makecell{Disconnected\\$\{1,1\}$ ZZ-instantons}  
\end{tabular}
\caption{The $\{1,1,1\}$ ZZ-instanton contribution with subtraction of disconnected instanton diagrams. 
}
\label{fig:111disc}
\end{figure}

We begin with the case of three $(1,1)$ ZZ-instantons, located at time coordinates $x_1$, $x_2$, $x_3$. The worldsheet diagram at order $e^{-3/g_s}$ is again given by a pair of discs, each containing one closed string vertex operator, such that the boundaries of the discs lie on one or two out of the three instantons. Extra care must be taken in subtracting off the disconnected instanton diagrams so as to normalize the vaccum amplitude, shown schematically in Figure \ref{fig:111disc}.

The contribution from a pair of discs ending on the same $(1,1)$ ZZ-instanton is
\ie
\hphantom{1} e^{-{3\over g_s}} \cN_1^3\frac{1}{3!} \int dx_1 dx_2&dx_3\, 3~\tikz[baseline={([yshift=-1.6ex]current bounding box.center)}]{\pic{disk1={1}}}~\tikz[baseline={([yshift=-1.6ex]current bounding box.center)}]{\pic{disk1={1}}} \,\Bigg[ \exp \Big(2~\tikz[baseline={([yshift=-1.6ex]current bounding box.center)}]{\pic{cyl11={1}{2}}}+2~\tikz[baseline={([yshift=-1.6ex]current bounding box.center)}]{\pic{cyl11={1}{3}}}+2~\tikz[baseline={([yshift=-1.6ex]current bounding box.center)}]{\pic{cyl11={2}{3}}}\Big) 
\\
-&  \exp \Big(2~\tikz[baseline={([yshift=-1.6ex]current bounding box.center)}]{\pic{cyl11={1}{2}}}\Big) -\exp \Big(2~\tikz[baseline={([yshift=-1.6ex]current bounding box.center)}]{\pic{cyl11={1}{3}}}\Big) -\exp \Big(2~\tikz[baseline={([yshift=-1.6ex]current bounding box.center)}]{\pic{cyl11={2}{3}}}\Big) +2\Bigg].
\label{oooxa}
\fe
The first two subtractions are due to the diagram with a disconnected instanton of type $\{1\}$, whereas the third subtraction is due to a disconnected instanton of type $\{1,1\}$. The last term in the bracket takes care of the over-subtraction of diagrams with two disconnected instantons of type $\{1\}$.


The contribution from a pair of discs ending on two separate $(1,1)$ ZZ-instantons is computed by
\ie{}
e^{-{3\over g_s}} \cN_1^3\frac{1}{3!} \int dx_1 dx_2&dx_3\, 3~\Bigg(\tikz[baseline={([yshift=-1.6ex]current bounding box.center)}]{\pic{disk1={1}}}~\tikz[baseline={([yshift=-1.6ex]current bounding box.center)}]{\pic{disk1={2}}}+\tikz[baseline={([yshift=-1.6ex]current bounding box.center)}]{\pic{disk1={2}}}~\tikz[baseline={([yshift=-1.6ex]current bounding box.center)}]{\pic{disk1={1}}}\Bigg) \exp \Big(2~\tikz[baseline={([yshift=-1.6ex]current bounding box.center)}]{\pic{cyl11={1}{2}}}\Big)\Bigg[ \exp \Big(2~\tikz[baseline={([yshift=-1.6ex]current bounding box.center)}]{\pic{cyl11={1}{3}}}+2~\tikz[baseline={([yshift=-1.6ex]current bounding box.center)}]{\pic{cyl11={2}{3}}}\Big)-1\Bigg].
\label{oooxb}
\fe
Here the subtraction involves only a single disconnected $\{1\}$ instanton.

After evaluating the integrals (along the Lorentzian time contour), (\ref{oooxa}) and (\ref{oooxb}) together give
\ie
e^{-\frac{3}{g_s}} 2\pi \delta(\omega_1-\omega_2) \cN_1^3 {64\pi^4\over 3}\sinh^2(\pi\omega_1)(1+e^{-2\pi\omega_1} + e^{-4\pi \omega_1}).
\label{eq:111zz}
\fe

\subsubsection{$\{2,1\}$ ZZ-instantons}
\label{sec:21zz}

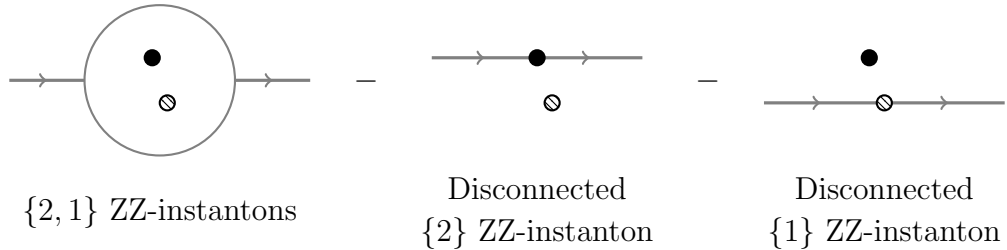
\begin{figure}[h!]
\centering
\begin{tabular}{c c c c c}
\begin{tikzpicture}
[yscale=1,line width=.8pt,
    decoration={
      markings,
      mark=at position 0.5 with {\arrow[line width=1pt]{>}}
    }
]
\draw[color=gray, thick] (0,0) circle [radius=1];
\draw[color=gray,very thick,postaction=decorate] (-2,0) -- (-1,0);
\draw[color=gray,very thick,postaction=decorate] (1,0) -- (2,0);
\draw[fill] (-0.1,0.3) circle [radius=0.1];
\draw[pattern=north west lines, pattern color=black] (0.1,-.3) circle [radius=0.1];
\end{tikzpicture} &
\begin{tikzpicture}
\draw[color=white] (0,-1)--(0,1);
\node at (0,0) {$-$};
\end{tikzpicture} &
\begin{tikzpicture}
[yscale=1,line width=.8pt,
    decoration={
      markings,
      mark=at position 0.5 with {\arrow[line width=1pt]{>}}
    }
]
\draw[color=white, thick] (0,0) circle [radius=1];
\draw[color=gray,very thick,postaction=decorate] (-1.5,0.3) -- (-0.1,0.3);
\draw[color=gray,very thick,postaction=decorate] (-0.1,0.3) -- (-0.2+1.5,0.3);
\draw[fill] (-0.1,0.3) circle [radius=0.1];
\draw[pattern=north west lines, pattern color=black] (0.1,-.3) circle [radius=0.1];
\end{tikzpicture} &
\begin{tikzpicture}
\draw[color=white] (0,-1)--(0,1);
\node at (0,0) {$-$};
\end{tikzpicture} &
\begin{tikzpicture}
[yscale=1,line width=.8pt,
    decoration={
      markings,
      mark=at position 0.5 with {\arrow[line width=1pt]{>}}
    }
]
\draw[color=white, thick] (0,0) circle [radius=1];
\draw[color=gray,very thick,postaction=decorate] (-1.5,-0.3) -- (0,-0.3);
\draw[color=gray,very thick,postaction=decorate] (0.2,-0.3) -- (1.7,-0.3);
\draw[fill] (-0.1,0.3) circle [radius=0.1];
\draw[pattern=north west lines, pattern color=black] (0.1,-.3) circle [radius=0.1];
\end{tikzpicture}
\\
\makecell{$\{2,1\}$ ZZ-instantons} & & 
\makecell{Disconnected\\$\{2\}$ ZZ-instanton} & & 
\makecell{Disconnected\\$\{1\}$ ZZ-instanton}
\end{tabular}
\caption{The $\{2, 1\}$ ZZ-instanton contribution with subtraction of disconnected instanton diagrams.}
\label{fig:12disc}
\end{figure}

Next, we consider a $(2,1)$ ZZ-instanton at time $x_1$ and a $(1,1)$ ZZ-instanton at time $x_2$. The measure factor is computed by the cylinder diagram between these two boundary conditions, as in (\ref{cyl11andn1}). We should also subtract off diagrams with a disconnected instanton, either of $(2,1)$ or $(1,1)$ type, as shown in Figure \ref{fig:12disc}.

The contribution from a pair of discs ending on the same ZZ-instanton is given by
\ie{}
& e^{-{3\over g_s}} \cN_1 \cN_2 \int dx_1 dx_2\, \left[ \tikz[baseline={([yshift=-1.6ex]current bounding box.center)}]{\pic{diskn={1}}}~\tikz[baseline={([yshift=-1.6ex]current bounding box.center)}]{\pic{diskn={1}}} \Bigg( \exp \Big(2~\tikz[baseline={([yshift=-1.6ex]current bounding box.center)}]{\pic{cyln1={1}{2}}}\Big)\right.  - 1 \Bigg) + \tikz[baseline={([yshift=-1.6ex]current bounding box.center)}]{\pic{disk1={2}}}~\tikz[baseline={([yshift=-1.6ex]current bounding box.center)}]{\pic{disk1={2}}} \Bigg( \exp \Big(2~\tikz[baseline={([yshift=-1.6ex]current bounding box.center)}]{\pic{cyln1={1}{2}}}\Big) - 1 \Bigg)
\\
&=e^{-\frac{3}{g_s}} 8\pi \delta(\omega_1-\omega_2) {\cal N}_1 {\cal N}_2 \int d\Delta x\left(\sinh^2(\pi\omega_1) +\sinh^2(2\pi\omega_1)\right)\left[\frac{(\Delta x)^2+\pi^2}{(\Delta x)^2+9\pi^2}-1\right],
\label{eq:(2,1)+(1,1)}
\fe
where the contribution from a pair of discs ending on the two different ZZ-instantons is
\ie{}
& e^{-{3\over g_s}} \cN_1 \cN_2 \int dx_1 dx_2 ~\left( \tikz[baseline={([yshift=-1.6ex]current bounding box.center)}]{\pic{diskn={1}}}~\tikz[baseline={([yshift=-1.6ex]current bounding box.center)}]{\pic{disk1={2}}} +\tikz[baseline={([yshift=-1.6ex]current bounding box.center)}]{\pic{disk1={2}}}~\tikz[baseline={([yshift=-1.6ex]current bounding box.center)}]{\pic{diskn={1}}} \right) \exp \Big(2~\tikz[baseline={([yshift=-1.6ex]current bounding box.center)}]{\pic{cyln1={1}{2}}}\Big)
\\
&= e^{-\frac{3}{g_s}}8\pi\delta(\omega_1-\omega_2) {\cal N}_1 {\cal N}_2 \int d\Delta x \, \frac{(\Delta x)^2+\pi^2}{(\Delta x)^2+9\pi^2}\, 2\cos(\omega_1\Delta x)\sinh(\pi\omega_1)\sinh(2\pi\omega_1) .
\fe
After evaluating the $\Delta x$-integral, we find the total contribution from $\{2,1\}$ ZZ-instanton configuration to be
\ie{}
& -e^{-\frac{3}{g_s}} 2\pi\delta(\omega_1-\omega_2) \cN_1 \cN_2 \frac{32\pi^2}{3}\sinh^2(\pi\omega_1) \left( e^{2\pi\omega_1} + 3 + 3 e^{-2\pi\omega_1} + 2 e^{-4\pi\omega_1} \right).
\label{eq:12zz}
\fe

\subsubsection{$\{3\}$ ZZ-instanton}

Finally, the contribution from a single $(3,1)$ ZZ-instanton at time $x$ is given by
\ie{}
&e^{-{3\over g_s}}\cN_3 \int dx ~\tikz[baseline={([yshift=-1.6ex]current bounding box.center)}]{\pic{diskn={1}}}~ \tikz[baseline={([yshift=-1.6ex]current bounding box.center)}]{\pic{diskn={1}}}= e^{-\frac{3}{g_s}} 8\pi \delta(\omega_1-\omega_2) {\cal N}_3 \sinh^2(3\pi \omega_1), 
\label{eq:3zz}
\fe
where the normalization factor ${\cal N}_{3,1}$ is so far undetermined.

Combining (\ref{eq:111zz}), (\ref{eq:12zz}), and (\ref{eq:3zz}), the order $e^{-3/g_s}$ contribution to the $1\to 1$ closed string amplitude remarkably agrees with the matrix model result (\ref{eq:1to1MM}) provided that we make the identification
\ie
\cN_3 = -{5\over 192\pi^2}.
\label{N31}
\fe

\subsection{4-instanton corrections to the closed string $1\to 1$ amplitude}
\label{sec:4zz}

The final example we consider is the order $e^{-4/g_s}$ correction to the $1\to 1$ closed string amplitude. There are various ZZ-instanton configurations that could contribute: $\{1,1,1,1\}$, $\{2,1,1\}$, $\{3,1\}$, $\{2,2\}$, $\{4\}$, all of which involve ZZ-instantons of type $(m,1)$. In view of (\ref{eq:Smn}), one may further suspect that a single ZZ-instanton of type $(2,2)$ could contribute at this order (not to be confused with $\{2,2\}$, which means two ZZ-instantons of type $(2,1)$). 
We will find a remarkable agreement of the total result with the matrix model, provided a suitable choice of the measure normalization factor ${\cal N}_4$ for the $(4,1)$ ZZ-instanton, and surprisingly, if we assume that the $(2,2)$ ZZ-instanton does not contribute, i.e. ${\cal N}_{(2,2)}=0$.

\subsubsection{$\{1,1,1,1\}$ ZZ-instantons}

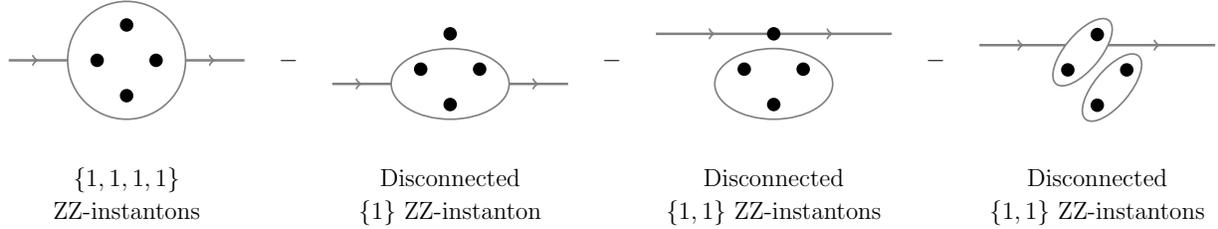
\begin{figure}[h!]
\centering
\resizebox{\columnwidth}{!}{%
\begin{tabular}{c c c c c c c}
\begin{tikzpicture}
[yscale=1,line width=.8pt,
    decoration={
      markings,
      mark=at position 0.5 with {\arrow[line width=1pt]{>}}
    }
]
\draw[color=gray, thick] (0,0) circle [radius=1];
\draw[color=gray,very thick,postaction=decorate] (-2,0) -- (-1,0);
\draw[color=gray,very thick,postaction=decorate] (1,0) -- (2,0);
\draw[fill] (0,0.6) circle [radius=0.1];
\draw[fill] (-0.5,0) circle [radius=0.1];
\draw[fill] (0.5,0) circle [radius=0.1];
\draw[fill] (0,-0.6) circle [radius=0.1];
\end{tikzpicture} &
\begin{tikzpicture}
\draw[color=white] (0,-1)--(0,1);
\node at (0,0) {$-$};
\end{tikzpicture} &
\begin{tikzpicture}
[yscale=1,line width=.8pt,
    decoration={
      markings,
      mark=at position 0.5 with {\arrow[line width=1pt]{>}}
    }
]
\draw[fill] (0,0.6) circle [radius=0.1];
\draw[fill] (-0.5,0) circle [radius=0.1];
\draw[fill] (0.5,0) circle [radius=0.1];
\draw[fill] (0,-0.6) circle [radius=0.1];
\draw[color=gray] (0,-0.25) ellipse (1 and 0.6);
\draw[color=gray,very thick,postaction=decorate] (-2,-0.25) -- (-1,-0.25);
\draw[color=gray,very thick,postaction=decorate] (1,-0.25) -- (2,-0.25);
\end{tikzpicture} &
\begin{tikzpicture}
\draw[color=white] (0,-1)--(0,1);
\node at (0,0) {$-$};
\end{tikzpicture} &
\begin{tikzpicture}
[yscale=1,line width=.8pt,
    decoration={
      markings,
      mark=at position 0.5 with {\arrow[line width=1pt]{>}}
    }
]
\draw[color=gray,very thick,postaction=decorate] (-2,0.6) -- (0,0.6);
\draw[color=gray,very thick,postaction=decorate] (0,0.6) -- (2,0.6);
\draw[fill] (0,0.6) circle [radius=0.1];
\draw[fill] (-0.5,0) circle [radius=0.1];
\draw[fill] (0.5,0) circle [radius=0.1];
\draw[fill] (0,-0.6) circle [radius=0.1];
\draw[color=gray] (0,-0.25) ellipse (1 and 0.6);
\end{tikzpicture} & 
\begin{tikzpicture}
\draw[color=white] (0,-1)--(0,1);
\node at (0,0) {$-$};
\end{tikzpicture} &
\begin{tikzpicture}
[yscale=1,line width=.8pt,
    decoration={
      markings,
      mark=at position 0.5 with {\arrow[line width=1pt]{>}}
    }
]
\draw[color=gray,very thick,postaction=decorate] (-2,0.42) -- (-0.5,0.42);
\draw[color=gray,very thick,postaction=decorate] (0.12,0.42) -- (2,0.42);
\draw[color=gray, fill=white, rotate around={50.2:(-0.25,0.3)}] (-0.25,0.3) ellipse (0.7cm and 0.3cm);
\draw[color=gray,rotate around={50.2:(0.25,-0.3)}] (0.25,-0.3) ellipse (0.7cm and 0.3cm);
\draw[fill] (0,0.6) circle [radius=0.1];
\draw[fill] (-0.5,0) circle [radius=0.1];
\draw[fill] (0.5,0) circle [radius=0.1];
\draw[fill] (0,-0.6) circle [radius=0.1];
\end{tikzpicture} \\
\\
\makecell{$\{1,1,1,1\}$ \\ZZ-instantons} & & 
\makecell{Disconnected \\$\{1\}$ ZZ-instanton} & & 
\makecell{Disconnected \\$\{1,1\}$ ZZ-instantons} & &
\makecell{Disconnected \\$\{1,1\}$ ZZ-instantons}\\
\end{tabular}
}
\caption{The $\{1,1,1,1\}$ ZZ-instanton contribution with subtraction of disconnected diagrams.}
\label{fig:1111disc}
\end{figure}

We begin with four $(1,1)$ ZZ-instantons, located at times $x_1, x_2, x_3, x_4$. The worldsheet diagrams again involve a pair of discs, with boundaries ending on either one or two out of the four instantons. The subtraction of disconnected diagrams is summarized schematically in Figure \ref{fig:1111disc}.

The contribution to the $1\to 1$ amplitude is computed as
\ie
\hphantom{1}&e^{-{4\over g_s}}\cN_1^4 {1\over 4!}\int \prod_{i=1}^4 dx_i  \left\lbrace 4 ~ \tikz[baseline={([yshift=-1.6ex]current bounding box.center)}]{\pic{disk1={1}}}~\tikz[baseline={([yshift=-1.6ex]current bounding box.center)}]{\pic{disk1={1}}} 
\left[  
\exp\Big(2\sum_{1\leq i<j\leq 4}\tikz[baseline={([yshift=-1.6ex]current bounding box.center)}]{\pic{cyl11={i}{j}}} \Big)
\right. \right.\\ 
&~~~~~~ - 3 \left( \exp\Big(2~\tikz[baseline={([yshift=-1.6ex]current bounding box.center)}]{\pic{cyl11={1}{2}}} + 2~\tikz[baseline={([yshift=-1.6ex]current bounding box.center)}]{\pic{cyl11={1}{3}}} + 2~\tikz[baseline={([yshift=-1.6ex]current bounding box.center)}]{\pic{cyl11={2}{3}}} \Big) - \exp\Big(2~\tikz[baseline={([yshift=-1.6ex]current bounding box.center)}]{\pic{cyl11={1}{2}}} \Big) - \exp\Big(2~\tikz[baseline={([yshift=-1.6ex]current bounding box.center)}]{\pic{cyl11={1}{3}}} \Big) - \exp\Big(2~\tikz[baseline={([yshift=-1.6ex]current bounding box.center)}]{\pic{cyl11={2}{3}}} \Big) + 2 \right) \\
&~~~~~~ \left. - \exp\Big(2~\tikz[baseline={([yshift=-1.6ex]current bounding box.center)}]{\pic{cyl11={2}{3}}} + 2~\tikz[baseline={([yshift=-1.6ex]current bounding box.center)}]{\pic{cyl11={2}{4}}} + 2~\tikz[baseline={([yshift=-1.6ex]current bounding box.center)}]{\pic{cyl11={3}{4}}} \Big) - 3 \left( \exp\Big(2~\tikz[baseline={([yshift=-1.6ex]current bounding box.center)}]{\pic{cyl11={1}{2}}} \Big) - 1 \right)\exp\Big(2~\tikz[baseline={([yshift=-1.6ex]current bounding box.center)}]{\pic{cyl11={3}{4}}} \Big)  \right] \\
& + 6 \Bigg(\tikz[baseline={([yshift=-1.6ex]current bounding box.center)}]{\pic{disk1={1}}}~\tikz[baseline={([yshift=-1.6ex]current bounding box.center)}]{\pic{disk1={2}}}+\tikz[baseline={([yshift=-1.6ex]current bounding box.center)}]{\pic{disk1={2}}}~\tikz[baseline={([yshift=-1.6ex]current bounding box.center)}]{\pic{disk1={1}}}\Bigg) \left[ \exp\Big(2\sum_{1\leq i<j\leq 4}\tikz[baseline={([yshift=-1.6ex]current bounding box.center)}]{\pic{cyl11={i}{j}}}  \Big) \right. \\
&~~~~~~  \left.\left.  - 2\left( \exp\Big(2~\tikz[baseline={([yshift=-1.6ex]current bounding box.center)}]{\pic{cyl11={1}{3}}} + 2~\tikz[baseline={([yshift=-1.6ex]current bounding box.center)}]{\pic{cyl11={2}{3}}} \Big) - 1 \right)\exp\Big(2~\tikz[baseline={([yshift=-1.6ex]current bounding box.center)}]{\pic{cyl11={1}{2}}} \Big)  -\exp\Big(2~\tikz[baseline={([yshift=-1.6ex]current bounding box.center)}]{\pic{cyl11={1}{2}}} + 2~\tikz[baseline={([yshift=-1.6ex]current bounding box.center)}]{\pic{cyl11={3}{4}}} \Big) \right] \right\rbrace \\
& = -e^{-{4\over g_s}} 2\pi \delta(\omega_1 - \omega_2) \cN_1^4 64 \pi^6 \sinh^2(\pi\omega_1)\left( 1 + e^{-2\pi\omega_1} + e^{-4\pi\omega_1} + e^{-6\pi\omega_1} \right).
\label{eq:4(1,1)zz}
\fe

\subsubsection{$\{2,1,1\}$ ZZ-instantons}

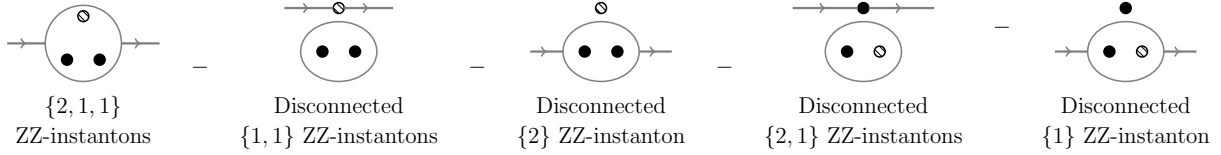
\begin{figure}[h!]
\centering
\resizebox{\columnwidth}{!}{
\begin{tabular}{c c c c c c c c c}
\begin{tikzpicture}
[yscale=1,line width=.8pt,
    decoration={
      markings,
      mark=at position 0.5 with {\arrow[line width=1pt]{>}}
    }
]
\draw[color=gray, thick] (0,0) circle [radius=0.7];
\draw[color=gray,very thick,postaction=decorate] (-1.4,0) -- (-0.7,0);
\draw[color=gray,very thick,postaction=decorate] (0.7,0) -- (1.4,0);
\draw[pattern=north west lines, pattern color=black] (0,.5) circle [radius=0.1];
\draw[fill] (-0.3,-0.3) circle [radius=0.1];
\draw[fill] (0.3,-0.3) circle [radius=0.1];
\end{tikzpicture}&
\begin{tikzpicture}
\node at (0,0) {$-$};
\end{tikzpicture} &
\begin{tikzpicture}
[yscale=1,line width=.8pt,
    decoration={
      markings,
      mark=at position 0.5 with {\arrow[line width=1pt]{>}}
    }
]
\draw[pattern=north west lines, pattern color=black] (0,.5) circle [radius=0.1];
\draw[fill] (-0.3,-0.30) circle [radius=0.1];
\draw[fill] (0.3,-0.3) circle [radius=0.1];
\draw[color=gray,very thick,postaction=decorate] (-1,0.5) -- (-0.1,0.5);
\draw[color=gray,very thick,postaction=decorate] (0.1,0.5) -- (1,0.5);
\draw[color=gray] (0,-0.3) ellipse (0.7 and 0.55);
\end{tikzpicture} &
\begin{tikzpicture}
\node at (0,0) {$-$};
\end{tikzpicture} &
\begin{tikzpicture}
[yscale=1,line width=.8pt,
    decoration={
      markings,
      mark=at position 0.5 with {\arrow[line width=1pt]{>}}
    }
]
\draw[pattern=north west lines, pattern color=black] (0,.5) circle [radius=0.1];
\draw[fill] (-0.3,-0.30) circle [radius=0.1];
\draw[fill] (0.3,-0.3) circle [radius=0.1];
\draw[color=gray] (0,-0.3) ellipse (0.7 and 0.55);
\draw[color=gray,very thick,postaction=decorate] (-1.3,-0.3) -- (-0.7,-0.3);
\draw[color=gray,very thick,postaction=decorate] (0.7,-0.3) -- (1.3,-0.3);
\end{tikzpicture} &
\begin{tikzpicture}
\node at (0,0) {$-$};
\end{tikzpicture} &
\begin{tikzpicture}
[yscale=1,line width=.8pt,
    decoration={
      markings,
      mark=at position 0.5 with {\arrow[line width=1pt]{>}}
    }
]
\draw[pattern=north west lines, pattern color=black] (0.3,-.3) circle [radius=0.1];
\draw[fill] (-0.3,-0.30) circle [radius=0.1];
\draw[fill] (0,0.5) circle [radius=0.1];
\draw[color=gray,very thick,postaction=decorate] (-1.3,0.5) -- (-0.1,0.5);
\draw[color=gray,very thick,postaction=decorate] (0.1,0.5) -- (1.3,0.5);
\draw[color=gray] (0,-0.3) ellipse (0.7 and 0.55);
\end{tikzpicture}&
\begin{tikzpicture}
\draw[color=white] (0,-1)--(0,1);
\node at (0,0) {$-$};
\end{tikzpicture} &
\begin{tikzpicture}
[yscale=1,line width=.8pt,
    decoration={
      markings,
      mark=at position 0.5 with {\arrow[line width=1pt]{>}}
    }
]
\draw[pattern=north west lines, pattern color=black] (0.3,-.3) circle [radius=0.1];
\draw[fill] (-0.3,-0.30) circle [radius=0.1];
\draw[fill] (0,0.5) circle [radius=0.1];
\draw[color=gray,very thick,postaction=decorate] (-1.3,-0.3) -- (-0.7,-0.3);
\draw[color=gray,very thick,postaction=decorate] (0.7,-0.3) -- (1.3,-0.3);
\draw[color=gray] (0,-0.3) ellipse (0.7 and 0.55);
\end{tikzpicture}
\\
\makecell{$\{2,1,1\}$\\ZZ-instantons} & & 
\makecell{Disconnected\\$\{1,1\}$ ZZ-instantons} & & 
\makecell{Disconnected\\$\{2\}$ ZZ-instanton}  & & 
\makecell{Disconnected\\$\{2,1\}$ ZZ-instantons}  & & 
\makecell{Disconnected\\$\{1\}$ ZZ-instanton}  
\end{tabular}
}
\caption{The $\{2,1,1\}$ ZZ-instanton contribution and subtraction of disconnected diagrams. }
\label{fig:112disc}
\end{figure}

Next we turn to the case of one $(2,1)$ ZZ-instanton and a pair of $(1,1)$ ZZ-instantons, located at times $x_1, x_2, x_3$ respectively. The subtraction of disconnected diagrams is summarized in Figure \ref{fig:112disc}.

The contribution to the amplitude is evaluated as
\ie{}
&e^{-{4\over g_s}} \cN_1^2 \cN_2 \frac{1}{2}\int dx_1 dx_2 dx_3 \left[\left(
\tikz[baseline={([yshift=-1.6ex]current bounding box.center)}]{\pic{disk1={1}}}~\tikz[baseline={([yshift=-1.6ex]current bounding box.center)}]{\pic{disk1={1}}} +
\tikz[baseline={([yshift=-1.6ex]current bounding box.center)}]{\pic{disk1={2}}}~\tikz[baseline={([yshift=-1.6ex]current bounding box.center)}]{\pic{disk1={2}}} +
\tikz[baseline={([yshift=-1.6ex]current bounding box.center)}]{\pic{diskn={3}}}~\tikz[baseline={([yshift=-1.6ex]current bounding box.center)}]{\pic{diskn={3}}}
\right)\right.
\\
&~~~~~ \times\left( \exp\Big(2~\tikz[baseline={([yshift=-1.6ex]current bounding box.center)}]{\pic{cyl11={1}{2}}} + 2~\tikz[baseline={([yshift=-1.6ex]current bounding box.center)}]{\pic{cyln1={3}{1}}} + 2~\tikz[baseline={([yshift=-1.6ex]current bounding box.center)}]{\pic{cyln1={3}{2}}}  \Big) - \exp\Big(2~\tikz[baseline={([yshift=-1.6ex]current bounding box.center)}]{\pic{cyl11={1}{2}}} \Big) - \exp\Big(2~\tikz[baseline={([yshift=-1.6ex]current bounding box.center)}]{\pic{cyln1={3}{1}}}\Big) - \exp\Big(2~\tikz[baseline={([yshift=-1.6ex]current bounding box.center)}]{\pic{cyln1={3}{2}}}\Big) + 2 \right)\\
& ~~~+ \left( \tikz[baseline={([yshift=-1.6ex]current bounding box.center)}]{\pic{disk1={1}}}~\tikz[baseline={([yshift=-1.6ex]current bounding box.center)}]{\pic{disk1={2}}} +
\tikz[baseline={([yshift=-1.6ex]current bounding box.center)}]{\pic{disk1={2}}}~\tikz[baseline={([yshift=-1.6ex]current bounding box.center)}]{\pic{disk1={1}}} \right)\exp\Big( 2~\tikz[baseline={([yshift=-1.6ex]current bounding box.center)}]{\pic{cyl11={1}{2}}} \Big) \left( \exp\Big( 2~\tikz[baseline={([yshift=-1.6ex]current bounding box.center)}]{\pic{cyln1={3}{1}}} + 2~\tikz[baseline={([yshift=-1.6ex]current bounding box.center)}]{\pic{cyln1={3}{2}}} \Big) - 1 \right) \\
&~~~ + \left. 2 \left( \tikz[baseline={([yshift=-1.6ex]current bounding box.center)}]{\pic{diskn={3}}}~\tikz[baseline={([yshift=-1.6ex]current bounding box.center)}]{\pic{disk1={1}}} +
\tikz[baseline={([yshift=-1.6ex]current bounding box.center)}]{\pic{disk1={1}}}~\tikz[baseline={([yshift=-1.6ex]current bounding box.center)}]{\pic{diskn={3}}} \right) \exp\Big( 2~\tikz[baseline={([yshift=-1.6ex]current bounding box.center)}]{\pic{cyln1={3}{1}}} \Big) \left( \exp\Big( 2~\tikz[baseline={([yshift=-1.6ex]current bounding box.center)}]{\pic{cyln1={3}{2}}} + 2~\tikz[baseline={([yshift=-1.6ex]current bounding box.center)}]{\pic{cyl11={1}{2}}} \Big) - 1 \right) \right] 
\\
& = e^{-{4\over g_s}} 2\pi \delta(\omega_1 - \omega_2) \cN_1^2\cN_2 32\pi^4 
\\
&~~~ \times \left[ \sinh^2(2\pi\omega_1)  +\sinh^2(\pi\omega_1)\left(2+e^{-2\pi\omega_1}+e^{-6\pi\omega_1} \right) +  2\sinh(\pi\omega_1)\sinh(2\pi\omega_1)\left( e^{-3\pi\omega_1}+e^{-5\pi\omega_1} \right) \right].
\label{eq:2(1,1)p1(2,1)zz}
\fe

\subsubsection{$\{3,1\}$ ZZ-instantons}

The contribution from the configuration of a $(3,1)$ together with $(1,1)$ ZZ-instanton is evaluated similarly to the case of section \ref{sec:21zz} as
\ie{}
& e^{-{4\over g_s}} \cN_1 \cN_3  \int dx_1 dx_2\, \left[ \tikz[baseline={([yshift=-1.6ex]current bounding box.center)}]{\pic{diskn={1}}}~\tikz[baseline={([yshift=-1.6ex]current bounding box.center)}]{\pic{diskn={1}}} \Bigg( \exp \Big(2~\tikz[baseline={([yshift=-1.6ex]current bounding box.center)}]{\pic{cyln1={1}{2}}}\Big)\right.  - 1 \Bigg) + \tikz[baseline={([yshift=-1.6ex]current bounding box.center)}]{\pic{disk1={2}}}~\tikz[baseline={([yshift=-1.6ex]current bounding box.center)}]{\pic{disk1={2}}} \Bigg( \exp \Big(2~\tikz[baseline={([yshift=-1.6ex]current bounding box.center)}]{\pic{cyln1={1}{2}}}\Big) - 1 \Bigg)
\\
& \left. ~~~~~~~~~~~~~~~~~~~~~~~~~~~~~~~~~~~~~~ + \left( \tikz[baseline={([yshift=-1.6ex]current bounding box.center)}]{\pic{disk1={2}}}~\tikz[baseline={([yshift=-1.6ex]current bounding box.center)}]{\pic{diskn={1}}} \,+\, \tikz[baseline={([yshift=-1.6ex]current bounding box.center)}]{\pic{diskn={1}}}~\tikz[baseline={([yshift=-1.6ex]current bounding box.center)}]{\pic{disk1={2}}} \right)\exp\Big(2~\tikz[baseline={([yshift=-1.6ex]current bounding box.center)}]{\pic{cyln1={1}{2}}}\Big) \right] \\
& = -e^{-{4\over g_s}} 2\pi \delta(\omega_1-\omega_2) \cN_1 \cN_3 12 \pi^2 \left[ \sinh^2(\pi\omega_1) + \sinh^2(3\pi\omega_1) + 2\sinh(\pi\omega_1)\sinh(3\pi\omega_1)e^{-4\pi\omega_1}\right].
\label{eq:1(1,1)p1(3,1)zz}
\fe

\subsubsection{$\{2,2\}$ ZZ-instantons}

The contribution from a pair of $(2,1)$ ZZ-instantons is evaluated similarly to the case of section \ref{sec:2zz} as
\ie
\hphantom{1}& e^{-{4\over g_s}} \cN_2^2\frac{1}{2}  \int dx_1 dx_2\, \left[ 2 ~ \tikz[baseline={([yshift=-1.6ex]current bounding box.center)}]{\pic{diskn={1}}}~\tikz[baseline={([yshift=-1.6ex]current bounding box.center)}]{\pic{diskn={1}}} \left( \exp\Big(2~\tikz[baseline={([yshift=-1.6ex]current bounding box.center)}]{\pic{cylnn={1}{2}}}\Big) - 1 \right) \right. \\
& \left. ~~~~~~~~~~~~~~~~~~~~~~~~~~~~~~~~~~~~~~ + \left( \tikz[baseline={([yshift=-1.6ex]current bounding box.center)}]{\pic{diskn={1}}}~\tikz[baseline={([yshift=-1.6ex]current bounding box.center)}]{\pic{diskn={2}}} \,+\, \tikz[baseline={([yshift=-1.6ex]current bounding box.center)}]{\pic{diskn={2}}}~\tikz[baseline={([yshift=-1.6ex]current bounding box.center)}]{\pic{diskn={1}}} \right)\exp\Big(2~\tikz[baseline={([yshift=-1.6ex]current bounding box.center)}]{\pic{cylnn={1}{2}}}\Big) \right] \\
& = -e^{-{4\over g_s}} 2\pi \delta(\omega_1-\omega_2) \cN_2^2 16 \pi^2 \sinh^2(2\pi\omega_1) \left( 1 + e^{-4\pi\omega_1} \right).
\label{eq:2(2,1)zz}
\fe

\subsubsection{$\{4\}$ ZZ-instanton and the putative $(2,2)$ ZZ-instanton}

Now we turn to new types of instantons that emerge at order $e^{-4/g_s}$. The contribution to the $1\to 1$ closed string amplitude from the ZZ-instanton configuration $\{4\}$, i.e. a single $(4,1)$ ZZ-instanton, is
\ie
e^{-{4\over g_s}}&\cN_4 \int dx_1 ~\tikz[baseline={([yshift=-1.6ex]current bounding box.center)}]{\pic{diskn={1}}}~ \tikz[baseline={([yshift=-1.6ex]current bounding box.center)}]{\pic{diskn={1}}}=e^{-{4\over g_s}} 2\pi \delta(\omega_1-\omega_2){\cal N}_{(4,1)} 4 \sinh^2(4\pi \omega_1),
\label{eq:1(4,1)zz}
\fe
where the normalization constant ${\cal N}_{(4,1)}$ is to be determined. A single $(2,2)$ ZZ-instanton, on the other hand, would give a contribution of the form
\ie
e^{-{4\over g_s}}&\cN_{(2,2)} \int dx_1 ~\tikz[baseline={([yshift=-1.6ex]current bounding box.center)}]{\pic{diskn={1}}}~ \tikz[baseline={([yshift=-1.6ex]current bounding box.center)}]{\pic{diskn={1}}}=e^{-{4\over g_s}} 2\pi \delta(\omega_1-\omega_2){\cal N}_{(2,2)} \frac{4\sinh^4(2\pi \omega_1)}{\sinh^2(\pi \omega_1)}.
\label{eq:1(2,2)zz}
\fe
Combining the results (\ref{eq:4(1,1)zz}), (\ref{eq:2(1,1)p1(2,1)zz}), (\ref{eq:1(1,1)p1(3,1)zz}), (\ref{eq:2(2,1)zz}), (\ref{eq:1(4,1)zz}), (\ref{eq:1(2,2)zz}), we find perfect agreement with the matrix model amplitude ${\cal A}_{1\to 1}^{4-{\rm inst},(0)}$ in (\ref{eq:1to1MM}) provided
\ie
\cN_4 = {35\over 2048\pi^2},
~~~~~ \cN_{(2,2)} = 0.
\label{ntwotwo}
\fe
The absence of $(2,2)$-type ZZ-instanton contribution leads us to suspect that in fact ${\cal N}_{(k,\ell)}=0$ whenever $k, \ell\geq 2$ (recall that $(k,\ell)$ and $(\ell,k)$ ZZ-boundary conditions are equivalent in $c=25$ Liouville theory), i.e. only the ZZ-instantons of type $(m,1)$ can contribute to closed string amplitudes in $c=1$ string theory. We will confirm this by extending the computation of the closed string $1\to 1$ amplitude to order $e^{-n/g_s}$ for all $n$.


\section{Closed string reflection amplitude to all instanton orders}
\label{sec:allorders}

In the worldsheet description, the order $e^{-n/g_s}$ contributions to the $1\to1$ amplitude of closed strings come from  all ZZ instanton configurations of type $\{m_1,...,m_\ell\}$, consisting of an $(m_i, 1)$ ZZ-instanton located at time $x_i$, for each $i=1,\cdots,\ell$, subject to $\sum_{i=1}^\ell m_i=n$. 

The worldsheet diagram with two discs whose boundaries lie on two different ZZ-instantons, say the ones at $x_1$ and $x_\ell$, is computed by
\ie{}
& \int_{\cal C} dx_1...dx_{\ell} \biggl( \tikz[baseline={([yshift=-1.6ex]current bounding box.center)}]{\pic{diskn={1}}}~\tikz[baseline={([yshift=-1.6ex]current bounding box.center)}]{\pic{diskn={\ell}}} \,+\, \tikz[baseline={([yshift=-1.6ex]current bounding box.center)}]{\pic{diskn={\ell}}}~\tikz[baseline={([yshift=-1.6ex]current bounding box.center)}]{\pic{diskn={1}}} \biggr) ~\exp\Big(2~\sum_{1\leq i<j\leq \ell}\tikz[baseline={([yshift=-1.6ex]current bounding box.center)}]{\pic{cylnn={i}{j}}}\Big) 
\label{eq:Lzz1to1}
\fe
where the integration contour ${\cal C}$ is taken along Lorentzian times $x_i$, for real energy $\omega$ of the closed string state. The cylinder diagram between an $(m_1,1)$ and an $(m_2,1)$ ZZ-instanton at Euclidean time separation $\Delta x^E$ evaluates to
\ie
\tikz[baseline={([yshift=-1.5ex]current bounding box.center)}]{\pic{cylnn={1}{2}}}=\frac{1}{2}\int_0^\infty \frac{dt}{t}\left(e^{\frac{\pi  t}{2}\left(m_1+m_2\right)^2 }-e^{\frac{\pi  t}{2}\left(m_1-m_2\right)^2 }\right)e^{-t\frac{(\Delta x^E)^2}{2\pi}}=\frac{1}{2}\ln\left[\frac{(\Delta x^E)^2-\pi^2(m_1-m_2)^2}{(\Delta x^E)^2 - \pi^2(m_1+m_2)^2}\right].
\label{cylm1mL}
\fe
Analytically continuing (\ref{cylm1mL}) to Lorentzian times, and using (\ref{eq:disk1pt}), we can write (\ref{eq:Lzz1to1}) explicitly as
\ie{}
& 4 \sinh(m_1 \pi \omega)\sinh(m_\ell \pi \omega) \int_{\cal C} dx_1...dx_{\ell} \left(e^{i\omega_1 x_1-i\omega_2 x_\ell}+e^{-i\omega_2 x_1+i\omega_1  x_\ell}\right)
\prod_{1\leq i<j\leq \ell}\frac{\left(x_i-x_j\right)^2+\pi^2\left(m_i-m_j\right)^2}{\left(x_i-x_j\right)^2+\pi^2\left(m_i+m_j\right)^2}
\\
&=16\pi \delta(\omega_1-\omega_2)\sinh(m_1 \pi \omega)\sinh(m_\ell \pi \omega)\int_{\cal C} dy_1 \cos(\omega_1 y_1)
\\
& \times\frac{y_1^2+\pi^2\left(m_1-m_\ell\right)^2}{y_1^2+\pi^2\left(m_1+m_\ell\right)^2}\int_{\cal C}dy_2...dy_{\ell-1}\prod_{2\leq i\leq \ell-1}\frac{y_i^2+\pi^2\left(m_i-m_\ell\right)^2}{y_i^2+\pi^2\left(m_i+m_\ell\right)^2}\prod_{1\leq i<j\leq \ell-1}\frac{\left(y_i-y_j\right)^2+\pi^2\left(m_i-m_j\right)^2}{\left(y_i-y_j\right)^2+\pi^2\left(m_i+m_j\right)^2},
\label{eq:Lzz1to1ints}
\fe
where we have defined $y_i\equiv x_i-x_\ell$ and integrated over $x_\ell$.

The integrals over $y_2,\cdots, y_{\ell-1}$ in (\ref{eq:Lzz1to1ints}) are linearly divergent at large $y_i$'s. As already seen in section \ref{sec:zzinst}, such divergences are cured by subtracting off disconnected diagrams that correspond to instanton corrections to the vacuum amplitude. In fact, we can conveniently take into account these subtractions by deforming the contour ${\cal C}$ to either $\mathbb{R} + i\infty$ or $\mathbb{R} + i\infty$ for various terms in the integrand, and simply keep the residue contributions while discarding the contour at infinity.

Let us first consider the integral over $y_{\ell-1}$. The poles in $y_{\ell-1}$ are located at 
\ie
y_{\ell-1} = y_i\pm i\pi (m_i+m_{\ell-1}),~~~1\leq i \leq \ell,~~~i\neq \ell-1,
\fe
where we defined $y_\ell\equiv 0$. After deforming the $y_{\ell-1}$ contour and discarding the contribution at infinity, we pick up the residue contribution, which now contains a set of poles in $y_{\ell-2}$ at
\ie
y_{\ell-2} = y_i\pm i\pi (m_i+m_{\ell-2})\pi,~~~y_i\pm i\pi (m_i+m_{\ell-2}+2m_{\ell-1}),~~~1\leq i \leq \ell,~~~i\neq \ell-1,\ell-2.
\fe
Note that some other potential poles in $y_{\ell-2}$ are canceled by zeroes in the numerator of the form $(y_{\ell-2}-y_i)^2+\pi^2(m_{\ell-2}-m_i)^2$.

We can iterate this procedure and integrate out $y_2,\cdots, y_{\ell-2}$. The remaining integrand in $y_1$ has poles at
\ie
y_1 = &\pm i\pi (m_1+m_\ell),
\\
& \pm i \pi (m_1+m_\ell+2m_{i_1}),
\\
& \pm i\pi (m_1+m_\ell+2m_{i_1}+2m_{i_2}),
\\
& \cdots
\\
& \pm i\pi (m_1+m_\ell+2m_{2}+...+2m_{\ell-1}), 
\label{polepos}
\fe
where the $i_k$'s are a set of distinct indices ranging from 2 to $\ell-1$. We will refer to the poles at $\pm i\pi (m_1+m_\ell)$ as the $0$-th kind, the poles at $\pm i\pi (m_1+m_\ell+2m_{i_1})$ as the first kind, the poles at  $\pm i\pi (m_1+m_\ell+2m_{i_1}+2m_{i_2})$ as the second kind, and so forth. The residue of the last line of (\ref{eq:Lzz1to1ints}) at any one of the poles of the $k$-th kind on the upper half $y_1$-plane appears to be given by the formula
\ie
Q_k = i (-1)^{\ell} \pi^{2\ell-3}2^{2\ell-4}k!(\ell-k-2)!\frac{m_1\cdots m_\ell}{n}.
\label{eq:kres}
\fe
While we have not proven (\ref{eq:kres}) in general, we have verified it explicitly for the case of $\{m_1,...,m_{\ell\leq 5}\}$ and $\{1,1,1,1,1,1\}$ instanton configurations. Using this, we can evaluate the $y_1$-integral as a sum over residues,
\ie
\int_{\cal C} dy_1\cos(\omega_1 y_1) \sum_{\{i_1,\cdots, i_k\}} {2p_{i_1,\cdots, i_k} Q_k\over y_1^2 - p_{i_1,\cdots,i_k}^2} = 2\pi i \sum_{\{i_1,\cdots, i_k\}} Q_k e^{i \omega_1 p_{i_1,\cdots, i_k}},
\label{eq:resint}
\fe
where $p_{i_1,\cdots,i_k} \equiv  i\pi (m_1+m_\ell+2m_{i_1}+\cdots + 2m_{i_k})$.

For an $\{m_1,...,m_\ell\}$ instanton configuration, we must sum over all possible assignment of boundary conditions for the worldsheet diagram, namely the $i$-th ZZ-instanton for the first disc and the $j$-th ZZ-instanton for the second disc, $i\not=j$, as well as a pair of discs whose boundaries lie on the same ZZ-instanton. So far we have not explicitly discussed the latter case. In fact, one can verify that (somewhat surprisingly) for a pair of discs ending on the first instanton (of type $m_1$), after suitable subtraction of disconnected instanton diagrams, the analogous moduli integral simply evaluates to (\ref{eq:resint}) with the replacement $\omega_1\to 0$. In this case one should also replace $m_\ell$ by $m_1$ in the prefactor of (\ref{eq:Lzz1to1ints}) and include an overall factor of ${1\over 2}$ (since two different diagrams are accounted for in (\ref{eq:Lzz1to1ints})).

Taking into account the prefactors in (\ref{eq:Lzz1to1ints}), we arrive at the following result for the $\{m_1,\cdots,m_\ell\}$ ZZ-instanton contribution to the reflection amplitude at order $e^{-n/g_s}$ ($n=\sum_{i=1}^\ell m_i$)
\ie
{\cal A}&^{n-\text{inst},(0)}_{1\to 1; \{m_1,\cdots,m_\ell\}}(\omega)=\frac{\cN_{m_1} \cdots \cN_{m_\ell}}{S}(-1)^{\ell-1}\pi^{2\ell-1}2^{2\ell+1}\frac{m_1\cdots m_\ell}{n} \Bigg[ (\ell-1)!\sum_{i=1}^\ell\sinh^2(\pi m_i\omega)
\\
&+ \sum_{1\leq i< j\leq\ell} 2\sinh(\pi m_i\omega)\sinh(\pi m_j\omega)\sum_{k=0}^{\ell-2}k!(\ell-2-k)!\sum_{S_{ij}^{(k)}}e^{-\pi\omega(m_i+m_j+2m(S_{ij}^{(k)}))}\Bigg].
\label{eq:lengthyboi}
\fe
Here $S$ is the symmetry factor of the ZZ-instanton configuration, defined as $S=\prod_a \ell_a!$ where $\ell_a$ is the number of $m_i$'s that are equal to $a$. The last sum in the second line is taken over all subsets $S_{ij}^{(k)}$ of $\{1,\cdots,\ell\}-\{i,j\}$ with $k$ elements. We have also defined $m(S_{ij}^{(k)})\equiv\sum_{q\in S_{ij}^{(k)}}m_q$.

The sum in the first line of (\ref{eq:lengthyboi}) represents the contribution from diagrams in which both discs end up on the same ZZ-instanton. The second line of (\ref{eq:lengthyboi}), coming from pairs of discs that end on different ZZ-instantons, can be simplified via the identity
\ie{}
& \sum_{1\leq i< j\leq\ell} 2\sinh(\pi m_i\omega)\sinh(\pi m_j\omega)\sum_{k=0}^{\ell-2}k!(\ell-2-k)!\sum_{S_{ij}^{(k)}}e^{-\pi\omega(m_i+m_j+2m(S_{ij}^{(k)}))}
\\
&=\frac{(\ell-1)!}{4}e^{-\pi \omega n}\left[\ell(e^{\pi\omega n}+e^{-\pi \omega n})-\sum_{i=1}^\ell
\left(e^{\pi\omega(n-2m_i)}+e^{-\pi\omega(n-2m_i)}\right)\right].
\fe
Using this and applying some simple rearrangements to (\ref{eq:lengthyboi}), we arrive at a compact expression for the full ZZ-instanton contribution to the closed string reflection amplitude at order $e^{-n/g_s}$,
\ie{}
&{\cal A}^{n-\text{inst},(0)}_{1\to 1}(\omega)
\\
&=\sum_{\{m_1,\cdots,m_\ell\}}\frac{\cN_{m_1}\cdots \cN_{m_\ell}}{S}(-1)^\ell 2^{2\ell}\pi^{2\ell-1}\frac{m_1\cdots m_\ell}{n}(\ell-1)! e^{-\pi\omega n}\sinh(\pi\omega n)
\left(\ell-\sum_{i=1}^\ell e^{2\pi\omega m_i}\right),
\label{eq:finalWS}
\fe
where the sum is taken over all (unordered) partitions $\{m_1,\cdots,m_\ell\}$ of the integer $n$.

Let us compare this with the matrix model result (\ref{aonegenform}) specialized to the $1\to 1$ amplitude (expanding out the hypergeometric function ${}_2F_1$) 
\ie
\left[ {\cal A}^{n-\text{inst},(0)}_{1\to 1}(\omega) \right]_{\rm MM} =\frac{2}{\pi}(-1)^{n+1}\frac{\sinh(\pi\omega n)}{n}e^{\pi\omega n}\sum_{k=0}^n\frac{(2k-3)!!}{(2k)!!}\frac{(2(n-k)-3)!!}{(2(n-k))!!}e^{-2\pi\omega k}.
\label{eq:finalMM}
\fe
The $\sinh(\pi \omega n)e^{\pi\omega n}$ term in (\ref{eq:finalWS}) comes from the $\{n\}$ ZZ-instanton only. Matching its coefficient against that of (\ref{eq:finalMM}), we fix $\cN_n$ to be
\ie
\cN_n=\frac{(-1)^{n}}{4\pi^2n}\frac{(2n-1)!!}{(2n)!!}.
\label{nnresult}
\fe
This agrees with the results of section \ref{sec:zzinst} explicitly computed for $n$ up to $4$. 

A term proportional to $\sinh(\pi \omega n) e^{\pi\omega (n-2k)}$ in (\ref{eq:finalWS}) comes from the sum over partitions $\{m_1,\cdots,m_\ell\}$ with at least one $m_i=n-k$, for $0\leq k\leq n-1$. We can reduce such a restricted sum to one that is over partitions $\{m_1,\cdots, m_{\ell-1}\}$ of the integer $k$. One can then verify that (\ref{eq:finalWS}) and (\ref{eq:finalMM}) are in complete agreement 
using (\ref{nnresult}) as well as the combinatorial identity
\ie
\sum_{\{m_1,\cdots,m_\ell\}} \frac{(-1)^\ell}{S}\ell!\prod_{i=1}^\ell \frac{(2m_i-1)!!}{(2m_i)!!}=-\frac{1}{n!}\frac{(2n-3)!!}{2^n},
\fe
which we have verified numerically. This also confirms our hypothesis that ZZ-instantons of type $(m, r)$ with $m,r\geq 2$ do not contribute, extending the result (\ref{ntwotwo}).

\section{Discussion}
\label{sec:discussion}

We have extended the analysis of \cite{Balthazar:2019rnh} to include the effects of multiple ZZ-instantons in $c=1$ string theory. Guided by a simple proposal of the non-perturbative matrix model dual \cite{Balthazar:2019rnh}, we presented a detailed prescription for computing multi-instanton contributions to closed string amplitudes from the worldsheet perspective. The ingredients can be summarized as follows.

The general D-instanton configuration that contributes to the closed string scattering involve $k$ ZZ-instantons of type $(m_1,1), \cdots, (m_k, 1)$, located at times $x_1, \cdots, x_k$. The integration measure in $x_k$ is computed by the partition function of open strings stretched between the ZZ-instantons, up to an overall normalization constant ${\cal N}_{m}$ for each type $(m,1)$, determined by comparison with the matrix model to be (\ref{nnresult}). The integration over the instanton moduli space is performed along the ``Lorentzian contour," namely over real Lorentzian time coordinates $x_1, \cdots, x_k$, so as to avoid the poles in Euclidean times. 

The worldsheet diagrams that contribute are those with boundaries that lie on the ZZ-instantons. The leading contribution at the $n$-instanton level comes from diagrams that involve multiple disconnected discs, each with one closed string vertex operator insertion, such that the boundaries of the discs reside on a subset of the ZZ-instantons. To compute subleading corrections in $g_s$, which has only been analyzed explicitly in the $n=1$ case in \cite{Balthazar:2019rnh}, would require the Fischler-Susskind-Polchinski mechanism for cancelation of divergences between worldsheet diagrams of different topologies \cite{Fischler:1986ci, Fischler:1986tb, Polchinski:1994fq}. Furthermore, to fix a finite constant ambiguity in the cancelation of divergence requires carefully dividing up the moduli space of punctured Riemann surfaces with boundaries using string field theory \cite{Sen:2019qqg}.

In this paper, we explicitly computed the leading $n$-instanton contributions to the $1\to k$ closed string amplitude for $n=2$, and to the $1\to 1$ closed string amplitude for all $n$. In these computations the FSP mechanism is not required, and the main subtlety has to do with the computation of the measure on the instanton moduli space, the choice integration contour, and the subtraction of disconnected diagrams in order to normalize the vacuum amplitude. In the end, we found striking agreement with the proposed matrix model dual.

One of the surprises uncovered by our computation is that, to correctly account for the non-perturbative corrections in the matrix model proposal of \cite{Balthazar:2019rnh}, we must take into account not only multiple $(1,1)$ ZZ-instantons, but also the $(m,1)$ ZZ-instantons with $m\geq 2$, even though the latter are constructed from non-unitary ZZ boundary conditions in the $c=25$ Liouville theory \cite{Zamolodchikov:2001ah}. On the other hand, our results suggest that the more general ZZ-instantons of type $(n,m)$ with $n,m\geq 2$ do not contribute to closed string amplitudes. It would be good to understand the reason behind this.

Another unusual feature of our computation is the choice of Lorentzian contour in the integration over the ZZ-instanton collective coordinates, namely their locations in time. This was partially motivated by the fact that the measure on the instanton moduli space has poles at Euclidean time separations, where open string tachyons stretched between ZZ-instantons become on-shell. In the computation of perturbative string amplitudes it is often useful to consider the analytic continuation to complex energies. If we Wick rotate the energies of the external closed string states to that of Euclidean signature, we can maintain the analyticity of instanton amplitudes by rotating the integration contour in the instanton collective coordinates toward a Euclidean one, provided that no poles are crossed. In other words, at imaginary energies we may equivalently work with the Euclidean contour, defined in a way that circumvents the poles (from either above or below, as dictated by continuity of the rotation from the Lorentzian contour).

The choice of instanton integration contour is tied to the breaking of time-reversal symmetry at the non-perturbative level. From the matrix model perspective, the proposed closed string vacuum state $|\Omega\rangle$ is such that the fermions occupy all $|E\rangle_R$ with $E\leq -\mu$ and none of the $|E\rangle_L$ states. This choice breaks time-reversal symmetry.\footnote{We thank Edward Witten for bringing this point to our attention.} This is also seen explicitly in that the instanton amplitudes do not obey the perturbative crossing symmetry relations \cite{Balthazar:2017mxh} upon analytically continuing $\omega\to -\omega$.

There are two important simplifications that are special to $c=1$ string theory underlying our analysis. The first is that the perturbative expansions of closed string amplitudes are Borel summable \cite{Balthazar:2019rnh} (assuming the perturbative duality with the matrix model). This renders the instanton corrections, on top of the Borel-resummed perturbative answer, unambiguously defined. The second is the simplicity of the moduli space of D-instantons, and in particular the absence of singularity in limits where multiple D-instantons collide. Neither of these features are expected to hold in, say, the ten-dimensional type IIB string theory. Nonetheless, we hope our analysis will pave the way toward understanding the effect of D-instantons more generally.

\section*{Acknowledgements}

We would like to thank Igor Klebanov, Juan Maldacena, Silviu Pufu, Nati Seiberg, and Edward Witten for discussions. BB thanks Princeton University, XY thanks University of Amsterdam, for hospitality during the course of this work. This work is supported in part by a Simons Investigator Award from the Simons Foundation, by the Simons Collaboration Grant on the Non-Perturbative Bootstrap, and by DOE grant DE-SC00007870. BB is supported by the Bolsa de Doutoramento FCT fellowship. VR is supported by the Harvard University Graduate Prize Fellowship.

\appendix

\bibliographystyle{JHEP}
\bibliography{MZZ}

\end{document}